\documentclass[12pt]{article}
\usepackage{epsf,epsfig,cite}
\usepackage{amssymb,amsmath,amsfonts}
\usepackage{graphicx}
\usepackage{epstopdf}

\newcount\timecount
\newcount\hours \newcount\minutes  \newcount\temp \newcount\pmhours

\hours = \time
\divide\hours by 60
\temp = \hours
\multiply\temp by 60
\minutes = \time
\advance\minutes by -\temp
\def\hour{\the\hours}
\def\minute{\ifnum\minutes<10 0\the\minutes
            \else\the\minutes\fi}
\def\clock{
\ifnum\hours=0 12:\minute\ AM
\else\ifnum\hours<12 \hour:\minute\ AM
      \else\ifnum\hours=12 12:\minute\ PM
            \else\ifnum\hours>12
                 \pmhours=\hours
                 \advance\pmhours by -12
                 \the\pmhours:\minute\ PM
                 \fi
            \fi
      \fi
\fi
}

\def\monthname{\relax\ifcase\month 0/\or January\or February\or
   March\or April\or May\or June\or July\or August\or September\or
   October\or November\or December\else\number\month/\fi}

\def\bold#1{\setbox0=\hbox{$#1$}%
     \kern-.025em\copy0\kern-\wd0
     \kern.05em\copy0\kern-\wd0
     \kern-.025em\raise.0433em\box0 }

\def\gappeq{\mathrel{\rlap {\raise.5ex\hbox{$>$}}
{\lower.5ex\hbox{$\sim$}}}}

\def\lappeq{\mathrel{\rlap{\raise.5ex\hbox{$<$}}
{\lower.5ex\hbox{$\sim$}}}}

\def\beq{\begin{equation}}
\def\eeq{\end{equation}}

\def\m12{m_{1\!/2}}

\input paperdef

\begin{document}
\begin{titlepage}
\pagestyle{empty}
\baselineskip=21pt

\begin{flushright}
ETHZ--IPP PR-2005-01
\end{flushright}

\vskip 0.05in
\begin{center}
{\large{\bf New allowed mSUGRA parameter space from variations of the trilinear scalar coupling $A_0$}}
\end{center}
\begin{center}
\vskip 0.05in
{\bf Luisa Sabrina Stark}$^1$, {\bf Petra H\"afliger}$^{1,2}$, {\bf Adrian Biland}$^1$ and {\bf Felicitas Pauss}$^1$\\
\vskip 0.05in
{\it
$^1${Institute for Particle Physics, ETH Z\"urich, CH-8093 Z\"urich, Switzerland}\\
$^2${Paul Scherrer Institute, CH-5232 Villigen PSI, Switzerland}\\
}
\vskip 1.5in
{\bf Abstract}
\end{center}
\baselineskip=18pt 
\noindent
In minimal Supergravity (mSUGRA) models the lightest supersymmetric particle (assumed to be the lightest neutralino $\chi^0_1$) provides an excellent cold dark matter (CDM) candidate. The supersymmetric parameter space is significantly reduced, if the limits on the CDM relic density $\Omega_{CDM}h^2$, obtained from WMAP data, are used.  Assuming a vanishing trilinear scalar coupling $A_0$ and fixed values of tan$\beta$, these limits result in narrow lines of allowed regions in the $m_0 - m_{1/2}$ plane, the so called WMAP strips. In this analysis the trilinear coupling $A_0$ has been varied within $\pm 4$~TeV. A fixed non vanishing $A_0$ value leads to a shift of the WMAP strips in the $m_0 - m_{1/2}$ plane.

\end{titlepage}
\baselineskip=18pt

\renewcommand{\thefootnote}{\alph{footnote}}
\setcounter{footnote}{0}

\section{Introduction}

The supersymmetric (SUSY\footnote{In the following the shortcut SUSY is used for supersymmetric as well as Supersymmetry.}) parameter space in minimal Supergravity (mSUGRA)
scenarios is usually studied in terms of the common scalar mass $m_0$,
the common gaugino mass $m_{1/2}$, the ratio of the Higgs expectation
values tan$\beta$ and the sign of the Higgsino mass parameter
$\mu$. However, the fifth free parameter, the common trilinear scalar coupling
$A_0$, was usually set to zero. In recent studies, the impact of $A_0$ on the SUSY parameter space was recognised \cite{ellisA0}. \\
In the mSUGRA framework the lightest neutralino lends itself as an excellent cold
dark matter (CDM) candidate, thus providing a connection between
particle physics and astrophysics. The inclusion of cosmological
experimental data allows to significantly reduce the mSUGRA parameter
space. The satellite born detector
WMAP\footnote{\tt http://map.gsfc.nasa.gov} measured the abundance of CDM
in the Universe to be $0.094\!~<\!~\Omega_{CDM}
h^2\!~<\!~0.129$~(at $2\sigma$ C.L.) \cite{WMAP}. Under the assumption of $A_0$~=~0~TeV and fixed tan$\beta$, only some narrow
lines in the $m_0-m_{1/2}$ plane are left over as allowed regions after including WMAP data \cite{ellis_postW}.
In this analysis we have studied the effects of non vanishing couplings $A_0$ systematically and found
extended areas in the mSUGRA parameter space, which no longer can be excluded.
In the next section we give a short theoretical introduction to SUSY and mSUGRA
with particular emphasis on the role of the trilinear scalar coupling $A_0$. In section
\ref{relic} we describe the constraints from cosmological data on the
mSUGRA parameter space. The tools used for this work are briefly
discussed in section~\ref{mc}, including comparisons of the different
Monte Carlo (MC) programs. In section~\ref{scan} we present a scan
over the mSUGRA parameter space. Allowing $A_0$ to vary, the WMAP
strips are broadened to extended areas. A general parametrisation of these allowed regions
 is complicated. Thus, we fixed $A_0$ to several different values and constructed parametrisations for the resulting lines, which are presented in section~\ref{results}.


\section{Supersymmetry}
\label{susy}
The Standard Model of particle physics (SM) is in stupendous agreement with
experimental measurements. Why should it then be extended? The
SM encounters several theoretical problems, which cannot be solved without the
introduction of new physics.  {\it (i)} In the SM the electroweak (EW) symmetry has to be broken in order to generate the masses of the weak gauge
bosons. The Higgs sector has been introduced to mediate EW spontaneous symmetry breaking. However, the Higgs mechanism is not experimentally established yet. {\it (ii)} The couplings of the three gauge interactions do not unify at some high
energy scale, so that the SM cannot easily be included in a grand unified theory (GUT). {\it (iii)} Due to quadratically divergent contributions
to the Higgs boson mass,  the huge gap between the EW and the GUT scale requires the introduction of a fine
tuned mass counter term in order to establish an intermediate Higgs mass. This problem is known as the hierarchy problem  \cite{hierarchy}.
{\it (iv)} Furthermore the SM does not provide a candidate for CDM.\\
SUSY \cite{SUSY,MSSM} is one of the best motivated candidates for physics beyond the SM. It cannot solve all the problems of the SM
but all those listed before. New particles at the TeV scale modify the $\beta$-functions of the three gauge couplings such
that the latter meet at about $10^{16}$~GeV \cite{SUSY-GUT}. SUSY-GUTs generate the EW symmetry breaking dynamically, if the
top mass ranges between about 100 and 200~GeV \cite{dEWSB}, in agreement with the measured top mass of 172.7~$\pm$~2.9~GeV \cite{top}. By connecting fermions with bosons the hierarchy problem is solved. The
quadratic divergences are cancelled systematically order by order, if the corresponding couplings between SM and SUSY particles are identical \cite{hierarchy}. Fine tuning of the counter terms is not required, if the masses of the SUSY particles are not too large, i.e. of the
$\mathcal{O}$(TeV). The lightest supersymmetric particle (LSP), if stable,
provides a good candidate for CDM \cite{cdm}.
However, several problems of the SM remain unexplained in SUSY extensions of the SM as e.g. the masses of the fermions or the origin of three generations.


\subsection{MSSM}
\label{MSSM}
The minimal supersymmetric extension of the SM (MSSM) \cite{MSSM},
requires a doubling of the SM degrees of freedom (d.o.f.) including two complex Higgs
doublets $H_d=(H^0_d,H^-_d)$ and $H_u=(H^+_u,H^0_u)$ giving mass to the down-type and up-type fermions, respectively. For each
fermionic d.o.f. a
corresponding bosonic one exists and vice versa, as can be inferred from Tab.\ref{fig:sparticles}. Two complex Higgs doublets are
needed for the theory to be supersymmetric and free of anomalies. After spontaneous symmetry breaking five of the eight states remain as
physical particles: two neutral CP-even (scalar), one neutral CP-odd (pseudoscalar) and two charged Higgs bosons. The scalar
superpartners of the left/right-handed fermion components $\tilde{Q}_{L},\tilde{u}_R,\tilde{d}_R$ and $\tilde{L}_{L},\tilde{e}_R$  mix with each other yielding the mass eigenstates $\tilde{Q}_{1,2}$ and $\tilde{L}_{1,2}$, respectively \cite{MSSM}. Since the mixing angles are proportional to the masses of the ordinary fermions, mixing effects are only important for the third-generation
sfermions. The four neutralinos $\chi^0_{1\dots 4}$
are linear combinations of the SUSY partners of the neutral gauge bosons $\tilde{W}^0$, $\tilde{B}^0$ and the neutral
Higgsinos $\tilde{H}^0_{u,d}$, the superpartners of the neutral components of the two Higgs doublets. Analogously the charged winos
$\tilde{W}^\pm$ and the charged Higgsinos $\tilde{H}_u^+,~\tilde{H}_d^-$  build up the four charginos $\chi^\pm_{1,2}$ as mass eigenstates.

\renewcommand{\arraystretch}{1.2}
\begin{table}
\begin{tabular}{|l|l|l|l|}
\hline
\multicolumn{2}{|c|}{Gauge Multiplets} & \multicolumn{2}{|c|}{Chiral Multiplets} \\
\hline
J=1  &J=1/2  & J=1/2  & J=0\\
\hline
Gluon $g$  & Gluino $\tilde{g}$  & Quark $Q$ & Squark $\tilde{Q}_{1,2}$\\
\hline
$W^{\pm},~W^0$\hspace *{1cm} & Wino $\tilde{W}^{\pm},~\tilde{W}^0$ \hspace *{1cm} & Lepton $L$ & Slepton $\tilde{L}_{1,2}$\\
\hline
$B^0$  & Bino $\tilde{B}^0$  & Higgsino $\tilde{H}_d^0,\tilde{H}_d^-,\tilde{H}_u^+,\tilde{H}_u^0$\hspace*{.3cm} & Higgs $H_d,H_u$\hspace *{1cm}\\
\hline
\end{tabular}
\vskip.4cm
\caption[]{\label{fig:sparticles}  \it
Particle content of the MSSM. The gauge multiplets mediate the interactions while the chiral multiplets contain the matter content. J denotes the spin quantum number. }
\end{table}
\noindent
The R-parity \cite{rparity}, defined as $R= (-1)^{3B+2S+L}$ with $B=$~baryon number, $S=$~spin and
$L=$~lepton number is introduced as a new discrete symmetry, which distinguishes SM particles ($R=1$) and their SUSY partners ($R=-1$). In R-parity conserving models the SUSY particles
(sparticles) can only be produced/annihilated in pairs, so that the LSP is stable.


\subsection{mSUGRA}
\begin{figure}[t]
 \setlength{\unitlength}{1cm}
 \centering
 \begin{picture}(5,7.0)
  \put(-2,-.5){\epsfig{file=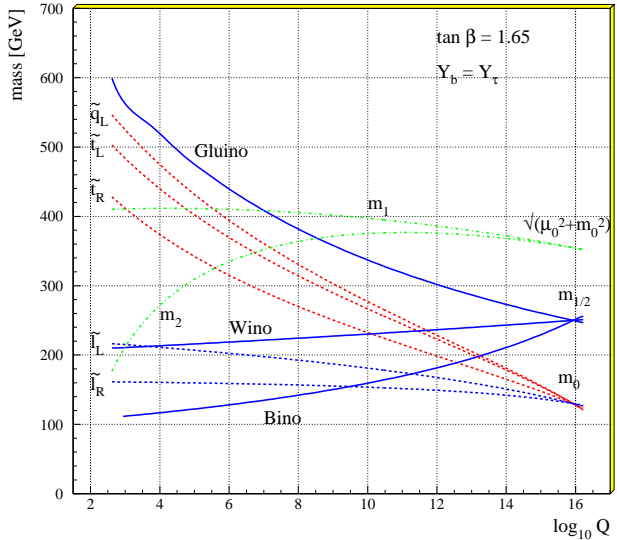,width=8.5cm}}
 \end{picture}
\caption[]{\it \label{fig:msugra} Unification of the sparticle masses at the GUT scale \cite{boer}. The common gaugino mass $m_{1/2}$ and the
common scalar mass $m_0$ are input parameters of mSUGRA.}
\end{figure}

Since no SUSY particle with the same mass as its SM partner has been discovered, SUSY has to be broken. Different scenarios for SUSY breaking mechanisms have been proposed. It is typically assumed, that the breaking takes place at a high energy scale. There are several models with different messenger particles (gravitons, gauge bosons, $\dots$) mediating the SUSY breaking effects down to the EW scale. We
concentrate on mSUGRA models with the graviton as the messenger particle \cite{msugra}. The key point of these models is the unification of the bosonic masses to the common scalar mass $m_0$, of the fermionic masses to the common gaugino mass $m_{1/2}$ and of the trilinear scalar couplings to $A_0$ at the GUT scale in addition to gauge coupling unification. As a consequence, the whole MSSM can be described by only five additional parameters to the SM ones (instead of more than 100 parameters as in the general MSSM): $m_0$, $m_{1/2}$, $A_0$, tan$\beta$ and sign$(\mu)$, the sign of the Higgsino mass parameter. All parameters of the MSSM can be derived by renormalisation group (RG) equations from the values of these five input parameters (which define a specific mSUGRA model) at the GUT scale, as illustrated in Fig.\ref{fig:msugra}.
\\
In R-parity conserving mSUGRA models the LSP is a neutralino in large regions of the parameter space, if cosmological bounds are included \cite{WMAP}. As the $\chi_1^0$ is electrically neutral, it does not directly couple to photons, an essential condition for any
dark matter candidate. 

\subsection{Trilinear scalar coupling $A_0$}
\begin{figure}[!t]
 \setlength{\unitlength}{1cm}
 \centering
 \begin{picture}(5,7.5)
  \put(-2,-.0){\epsfig{file=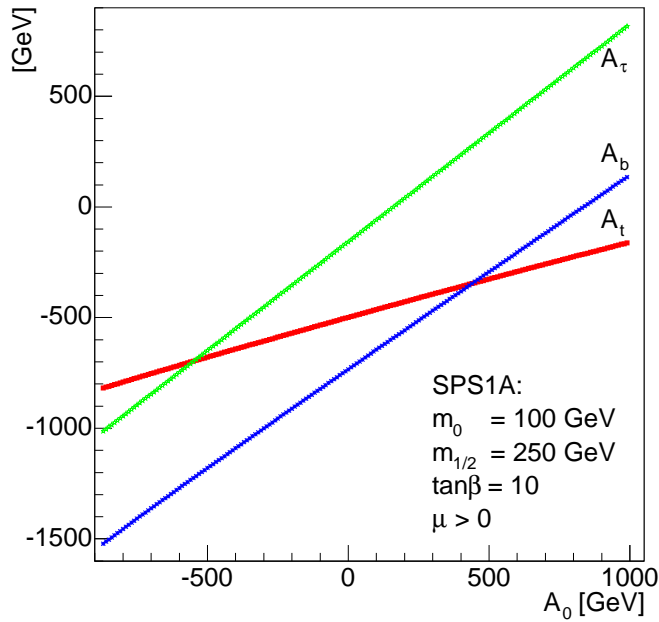,width=9.5cm}}
 \end{picture}
\caption[]{\it \label{fig:A} Dependence of the trilinear scalar couplings $A_{t},~A_{b}$ and $A_{\tau}$ at the EW scale on $A_0$ for
the Snowmass point SPS1A \cite{snow}. Values of $A_0$ outside the displayed region are excluded because of numerical problems of the RG equations. }
\end{figure}

In Fig.\ref{fig:A} the linear $A_0$ dependence of the trilinear couplings of the third generation, $A_{t,b,\tau}$, is depicted. The evolution of $A_0$ down to the EW scale determines the couplings $A_{u,d,l}$ for up-type, down-type squarks and charged sleptons resulting in the approximate relation:
\begin{eqnarray}
A_k = d_kA_0+d'_k\m12\hspace*{1cm}\mbox{with}\hspace*{1cm}k = u,d,l.
\end{eqnarray} 
 The coefficients $d_k$ depend on the corresponding Yukawa couplings and are of $\mathcal{O}(1)$, increasing for decreasing masses. As the top quark is much heavier than the bottom quark and the tau lepton, the slope of $A_t$ is smaller for small $\tan\beta$ (Fig.\ref{fig:A}). The coefficients $d'_k$ depend additionally on the gauge couplings and are of order unity \cite{zerwas}. The RG evolution of $A_{u,d,l}$ from the GUT to the EW scale generates the corresponding terms of the SUSY breaking part of the Lagrangian in the low energy limit:
\begin{eqnarray}
\mathcal{L}_{soft} = \frac{g}{\sqrt{2}M_W}\varepsilon_{ij}\left[
 \frac{M_u}{\sin\beta}A_uH^i_u\tilde{Q}^j_L\tilde{u}^\dagger_R
+\frac{M_d}{\cos\beta}A_dH^i_d\tilde{Q}^j_L\tilde{d}^\dagger_R
+\frac{M_l}{\cos\beta}A_lH^i_d\tilde{L}^j_L\tilde{e}^\dagger_R\right],
\label{eq:eq1}\end{eqnarray}
\vskip.1cm
\noindent
where $g$ denotes the coupling constant of SU(2)$_L$ and $\varepsilon_{ij}$ is the two-dimensional antisymmetric tensor. $M_W$ is the mass
of the W boson and $M_{u,d,l}$ are the masses of the up/down-type quarks and the charged leptons, respectively. These terms of the Lagrangian introduce interactions between MSSM Higgs bosons, ``left-'' and ``right-handed'' sfermions, in addition to the usual interaction mediated by the Yukawa couplings. These additional couplings are proportional to the
masses of the corresponding fermions (Eq.\ref{eq:eq1}), so that they are only relevant for the third generation. 

\section{Relic density}
\label{relic}

In the early Universe the interaction rate ($\Gamma$) of a species of particles must be larger than the expansion rate of the Universe. Otherwise the thermal equilibrium cannot be maintained and the particles decouple at the freeze out temperature $T_f$, i.e. if the following condition is fulfilled: 
\begin{eqnarray}
\Gamma=n~\!\langle\sigma v\rangle=H\hspace{1cm}\mbox{at} \hspace{.5cm}T_f,
\label{eq:freeze}
\end{eqnarray}
where $n$ denotes the number density of the particle species, $H$ is the Hubble constant and $\langle\sigma v\rangle$ is the thermal average of the total annihilation cross section times the velocity of the corresponding particles. The ratio of the number of relic particles to the total entropy in the Universe remains constant after they are  frozen out. Since the present entropy density is known, the present number of these particles can be approximately determined by using the freeze out condition (Eq.\ref{eq:freeze}). The relic density of a particle species $X$ can be estimated as \cite{jungman}:
\begin{equation}
        \Omega_X h^2 \approx \frac{3 \times 10^{-27}~cm^3s^{-1}}{\langle\sigma v\rangle_X},
\end{equation}
\noindent
where $h$ denotes the Hubble constant in units of 100~km~sec$^{-1}$~Mpc$^{-1}$. The cross section is proportional to the squared matrix element. The matrix element for a given process is inversely proportional to the squared mass of a heavy propagator particle. Assuming SUSY CDM, the propagator particle may also be a sparticle. In the case of neutralino annihilation, the cross section contains several different channels, thus the dependence of $\Omega_{\chi}h^2$ on the mSUGRA parameters is not trivial.
\begin{figure}
\begin{center}
\includegraphics[width=.55\textwidth]{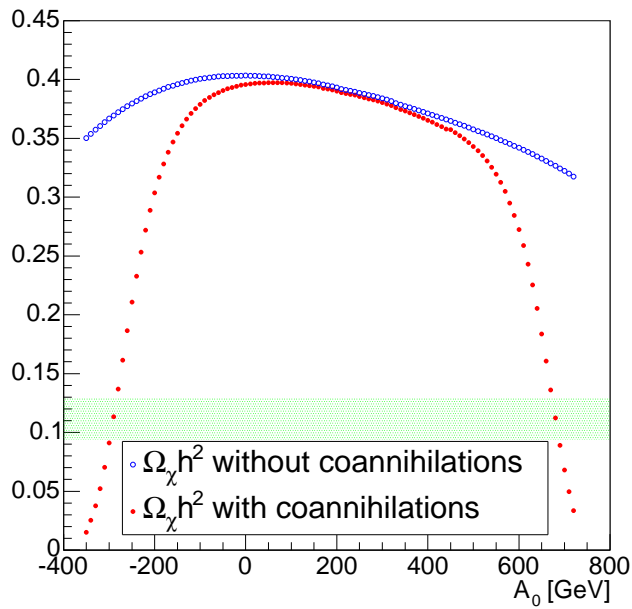}
\caption{\it Difference between the values of the relic density for calculations without (blue circles) and with (red points) coannihilation processes. The input parameters are taken from the Snowmass point SPS1B ($m_0$~=~200~GeV, $\m12$~=~400~GeV, tan$\beta$~=~30, $\mu>$~0) and $A_0$ has been varied within $\pm$1~TeV. The green shaded area shows the region allowed by the WMAP data.}
\label{coann}
\end{center}
\end{figure}

\noindent
Moreover, it is important to include all possible coannihilation processes \cite{edsjo} between the LSP and the next heavier sparticles as can be inferred from Fig.\ref{coann} for the Snowmass point SPS1B ($m_0$~=~200~GeV, $\m12$~=~400~GeV, tan$\beta$~=~30, $\mu>$~0)~\cite{snow}. $A_0$ has been varied within $\pm$1~TeV\footnote{Numerical problems of the RG equations occur for $A_0 <-$~350~GeV and $A_0 >$~720~GeV.}. Without coannihilations the Snowmass point SPS1B would be excluded by the WMAP data. Including coannihilations several models with $A_0 \ne 0$~TeV are allowed. The formulas including coannihilations are more involved~\cite{edsjo}.

\begin{figure}[!h]
\begin{center}
\hspace*{-.5cm}\includegraphics[width=0.52\textwidth]{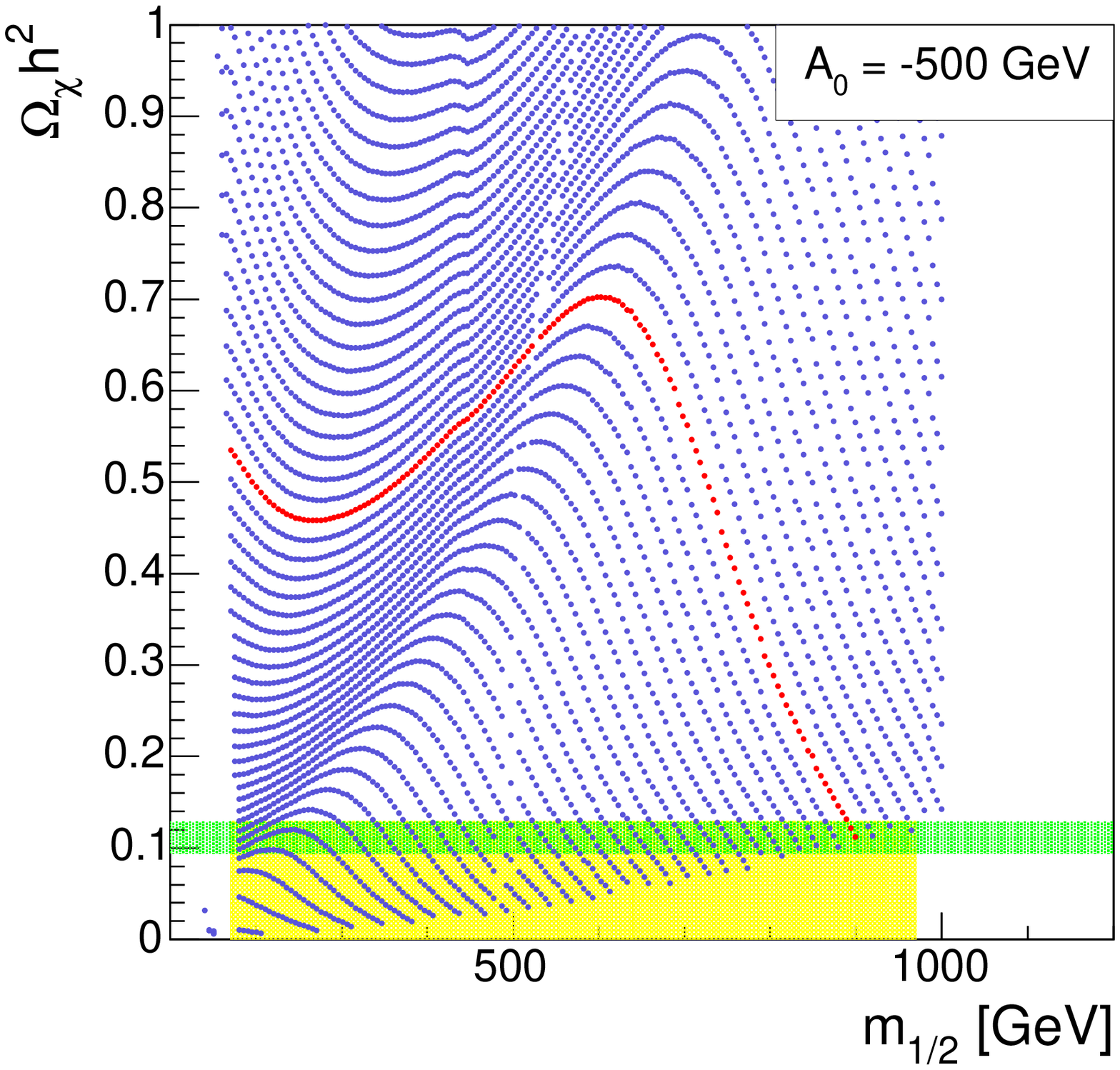}
\hspace*{-.5cm}\includegraphics[width=0.52\textwidth]{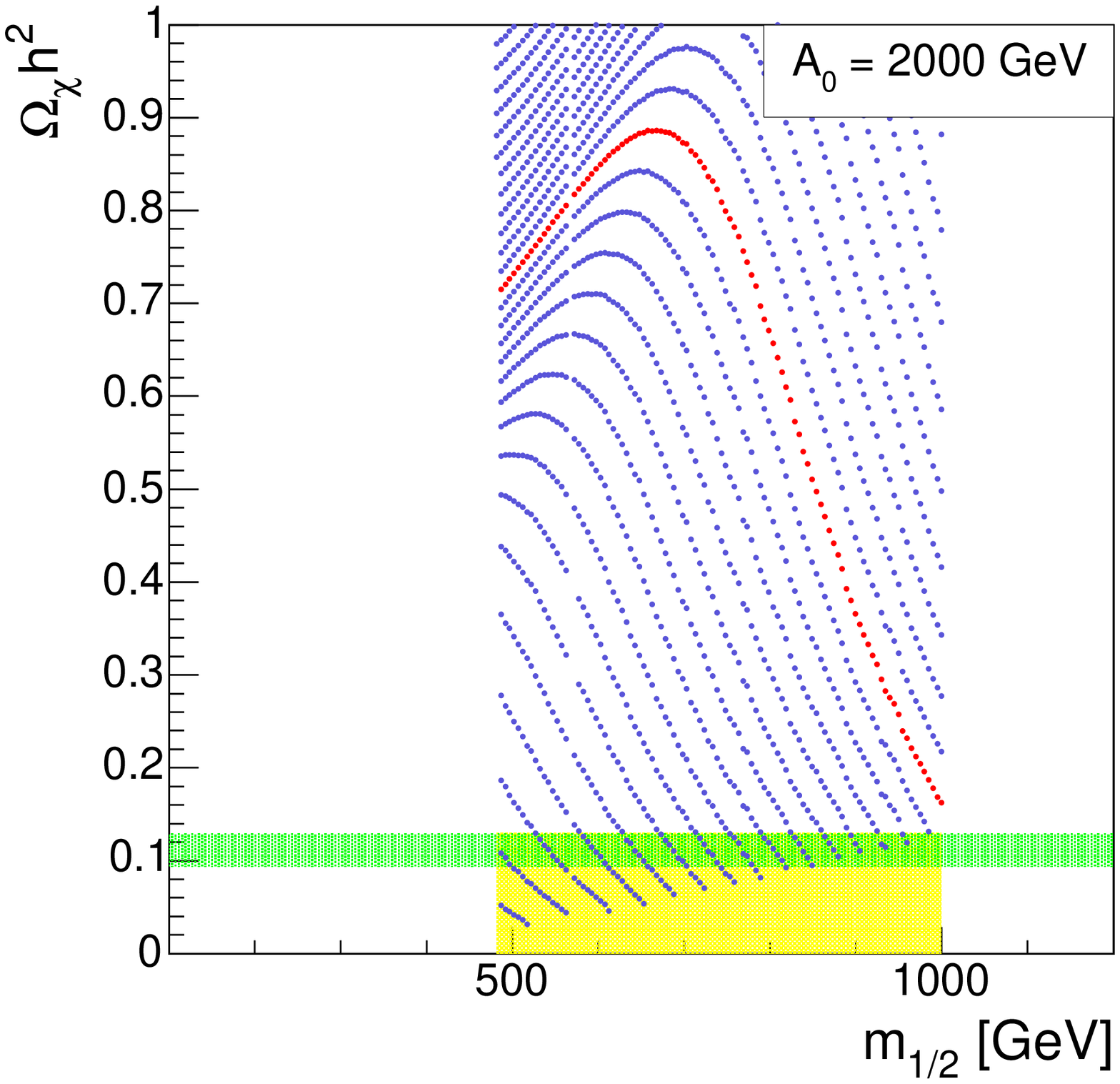}
\caption{\it The relic density $\Omega_\chi h^2$ as a function of $\m12$ obtained using the ISAJET and the DarkSUSY Monte Carlo programs, for tan$\beta$~=~10, $\mu >$~0, $\m12 \le$~1~TeV, $m_0$ varied between 0 and 300~GeV and $A_0$~=~$-$500~GeV (left) and $A_0$~=~2000~GeV (right). The red lines belong to $m_0$~=~190~GeV (left) and $m_0$~=~235~GeV (right). Each single point denotes one set of mSUGRA parameters. The green shaded area shows the region allowed by the WMAP data and the yellow shaded area indicates the resulting allowed $\m12$ region.}
\label{oh2_form}
\end{center}
\end{figure}
\noindent
The relic neutralino density (incl. coannihilation) for $\tan\beta$~=~10, $A_0=-$~500~GeV and +2000~GeV as a function of $\m12$ with 0~$<m_0<$~300~GeV is depicted in Fig.\ref{oh2_form}. The lines result from fixed values of $m_0$, the step sizes for $m_0,\m12$ are chosen to be 5~GeV for both plots. This is exemplified by the red lines, which correspond to $m_0$~=~190~GeV in the left plot  and to $m_0$~=~235~GeV in the right plot in Fig.\ref{oh2_form}. Values below the red lines belong to smaller $m_0$ values and lines above to larger ones. Models with $\m12\lesssim$~170~GeV in the left plot and with $\m12\lesssim$~480~GeV in the right plot are excluded by collider constraints. With growing $\m12$ the lower limit of the relic density becomes larger.

\begin{figure}[h]
\begin{center}
\hspace{-.8cm}\includegraphics[width=.499\textwidth]{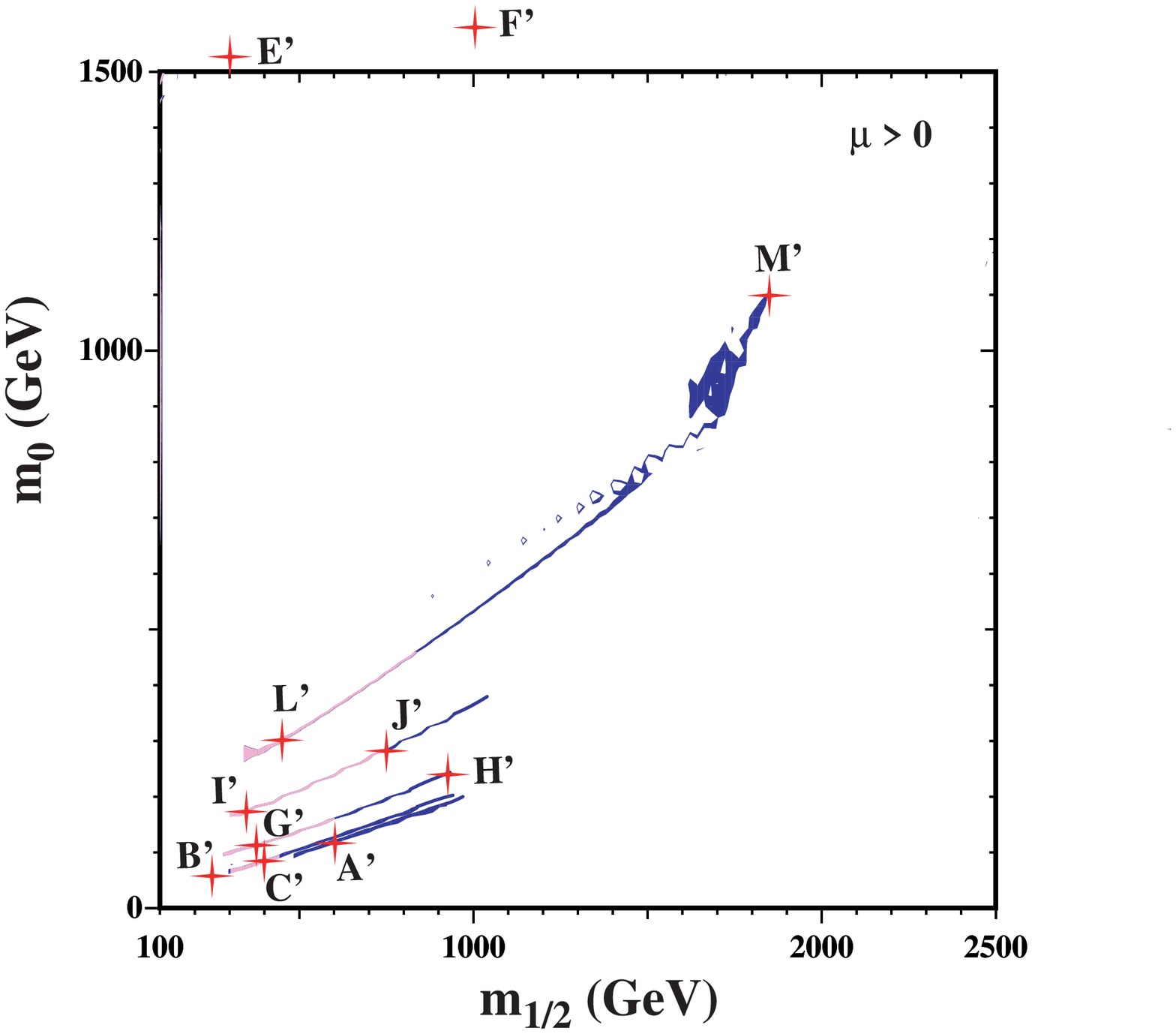}
\hspace{-.5cm}\includegraphics[width=.499\textwidth]{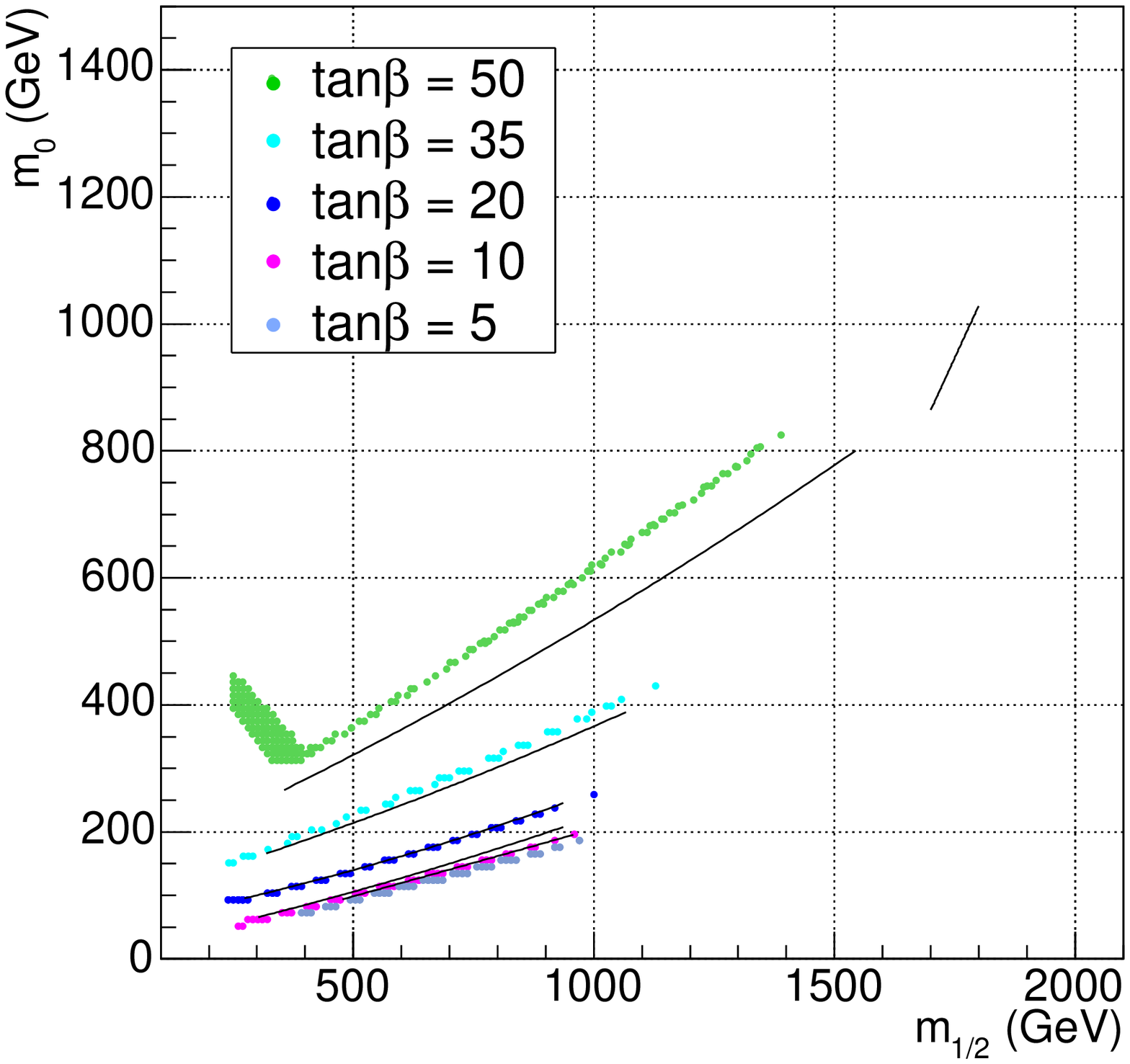}
\caption{\it Regions in the $m_0-\m12$ plane accounting for 0.094~$<\Omega_{\chi}h^2<$~0.129 for different tan$\beta$, $\mu >$~0 and $A_0 = 0$. Left plot: WMAP strips from Ref.\cite{ellis_postW} are shown with the post WMAP benchmark points (A'$\dots$M'). Right plot: The corresponding allowed regions obtained with ISAJET and DarkSUSY are shown. The black lines are the parametrisation given for the WMAP strips in the left plot.}
\vskip-.5cm
\label{Wstrips}
\end{center}
\end{figure}
\noindent
Assuming that CDM entirely consists of LSPs, the application of cosmological constraints on the mSUGRA parameter space is possible. The WMAP experiment constrained the relic CDM density to the narrow range 0.094~$<~\Omega_{CDM} h^2<$~0.129. For vanishing $A_0$, only some narrow strips remain as can be inferred from Fig.\ref{Wstrips}. The left plot has been obtained by the authors of Ref.\cite{ellis_postW} using their own Monte Carlo programs. The parametrisation of these data defines the WMAP strips (black lines in the right plot). The right plot displays the results of this study, using the ISAJET and DarkSUSY Monte Carlo generators, for comparison. For small values of $\tan\beta$ the agreement is good, but the differences increase with $\tan\beta$. The mSUGRA models with large values of $\tan\beta, m_0$ and $\m12$, shown in the left plot (in vicinity of M'), are excluded in the right plot since the relic densities are too large. The extended region for $\tan\beta$~=~50 and small values of $m_0,~\m12$ obtained with ISAJET does not show up in the left plot.
If we include variation of the parameter $A_0$ within $\pm$4~TeV, the WMAP strips are broadened to extended areas as will be shown in section \ref{scan}.


\section{Monte Carlo generators}
\label{mc}

As already stated in section \ref{susy}, the RG equations  have to be applied to evolve the mSUGRA parameters from the GUT scale down to the EW scale. For these evolutions two different Monte Carlo programs have been used:
\begin{enumerate}
\item {\bf SuSpect} \cite{suspect}:
SuSpect 2.2 is a Fortran code used to determine the SUSY particle spectra. The calculations are done in the MSSM framework assuming R-parity and CP conservation. It can also be applied to constrained scenarios as mSUGRA, anomaly mediated SUSY breaking (AMSB) and gauge mediated SUSY breaking (GMSB) models. The algorithm incorporates RG equations to evolve the parameters between the EW and the GUT scale complemented by constraints from radiative EW symmetry breaking. 
        
\item {\bf ISAJET} \cite{isajet}:   
the ISAJET 7.69 package contains ISASUSY, which treats the production and decays of supersymmetric particles. The calculations are done within the MSSM framework, if the input parameters are provided at the EW scale. The ISAJET package provides in addition the possibility to choose the input parameters at the GUT scale and then to perform the RG evolution down to the EW scale.
\end{enumerate}
\noindent
The generated SUSY spectra have been linked \cite{pepe} to the {\bf DarkSUSY} program to calculate the relic density.

\begin{enumerate}
\item[3.]{\bf DarkSUSY} \cite{darksusy1}:
DarkSUSY 4.00 is currently one of the most advanced programs to perform DM calculations in a SUSY framework.  It computes the relic density of the lightest neutralino which is assumed to be the DM particle. The calculations include the impacts of resonances, pair production thresholds and coannihilation processes \cite{edsjo}. To check the validity of the input parameters the presently known bounds from accelerators are included. This package also computes fluxes for a large variety of detector~types.
\end{enumerate}

\subsection{Differences between SuSpect and ISAJET}
\begin{figure}[!h]
\begin{center}
\hspace*{-.7cm}\includegraphics[width=.53\textwidth]{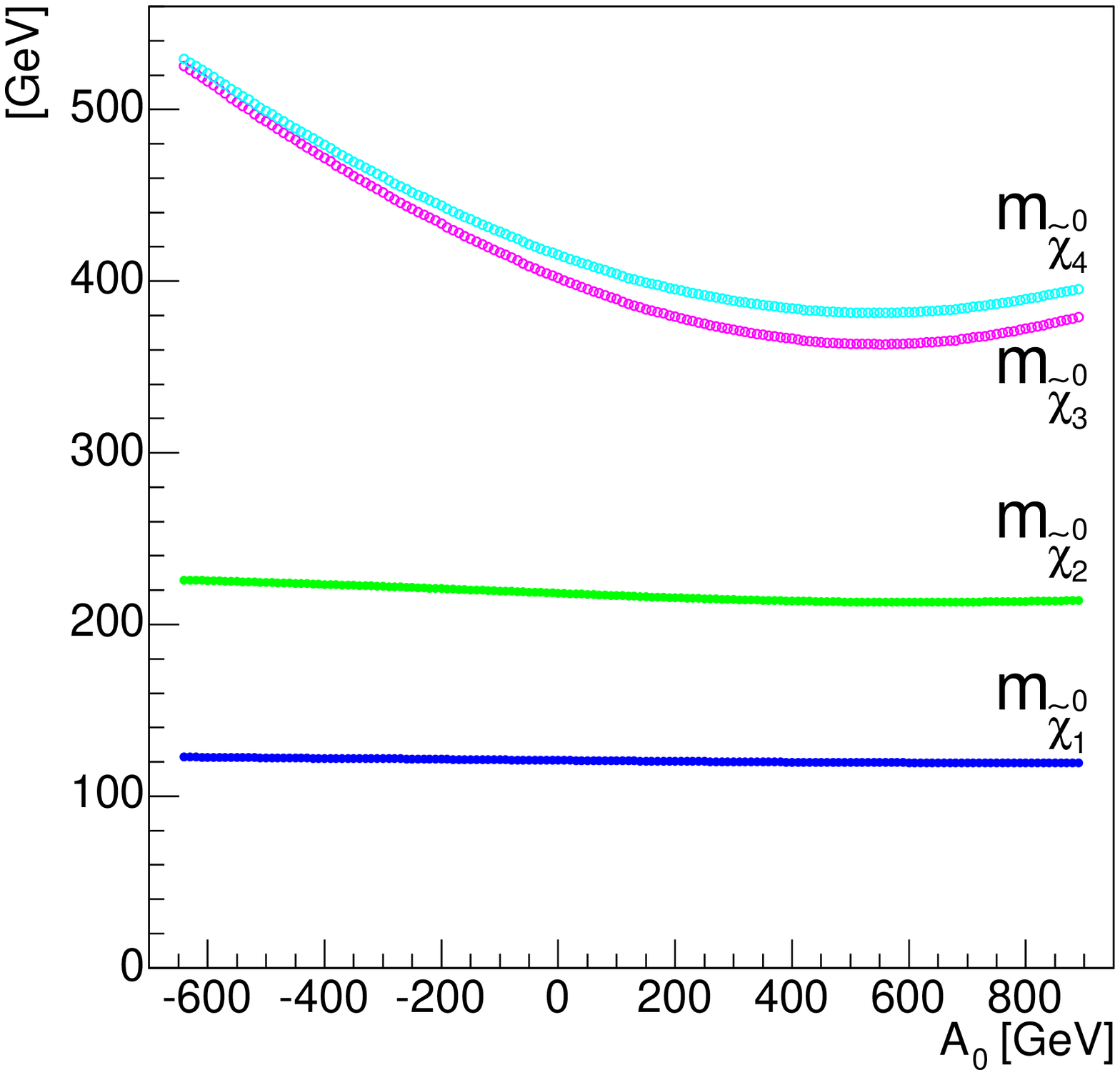}
\hspace*{-.7cm}\includegraphics[width=.53\textwidth]{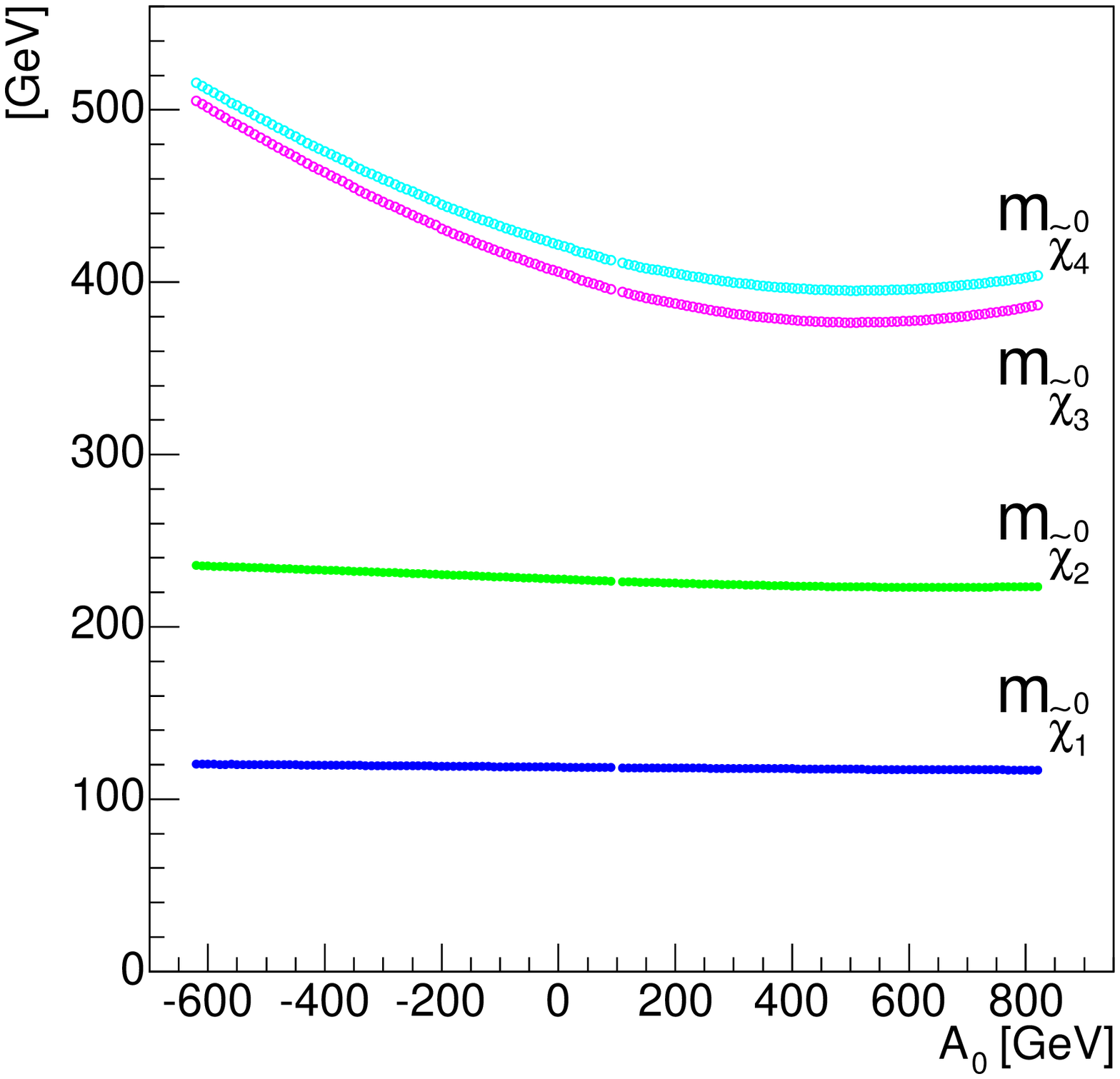}
\caption{\it Neutralino masses as a function of $A_0$ obtained with the SuSpect (left) and the ISAJET (right) programs for the Snowmass point SPS4 ($m_0$~=~400~GeV, $\m12$~=~300~GeV, tan$\beta$~=~50, $\mu >$~0).}
\label{SPS4_mchi}
\end{center}
\end{figure}

\begin{figure}[!h]
\vskip-.3cm
\begin{center}
\hspace*{-.3cm}\includegraphics[width=.50\textwidth]{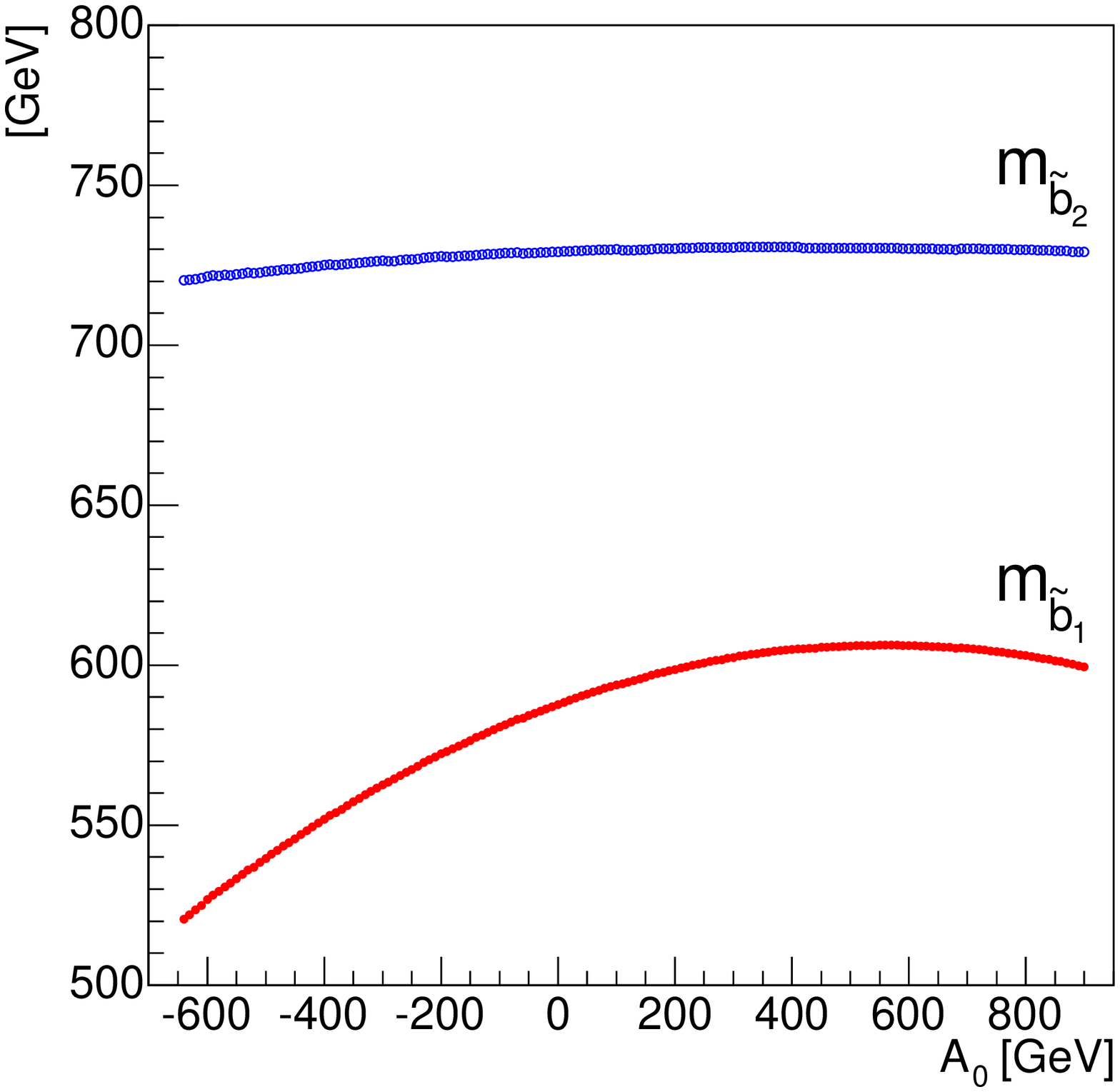}
\hspace*{-.3cm}\includegraphics[width=.50\textwidth]{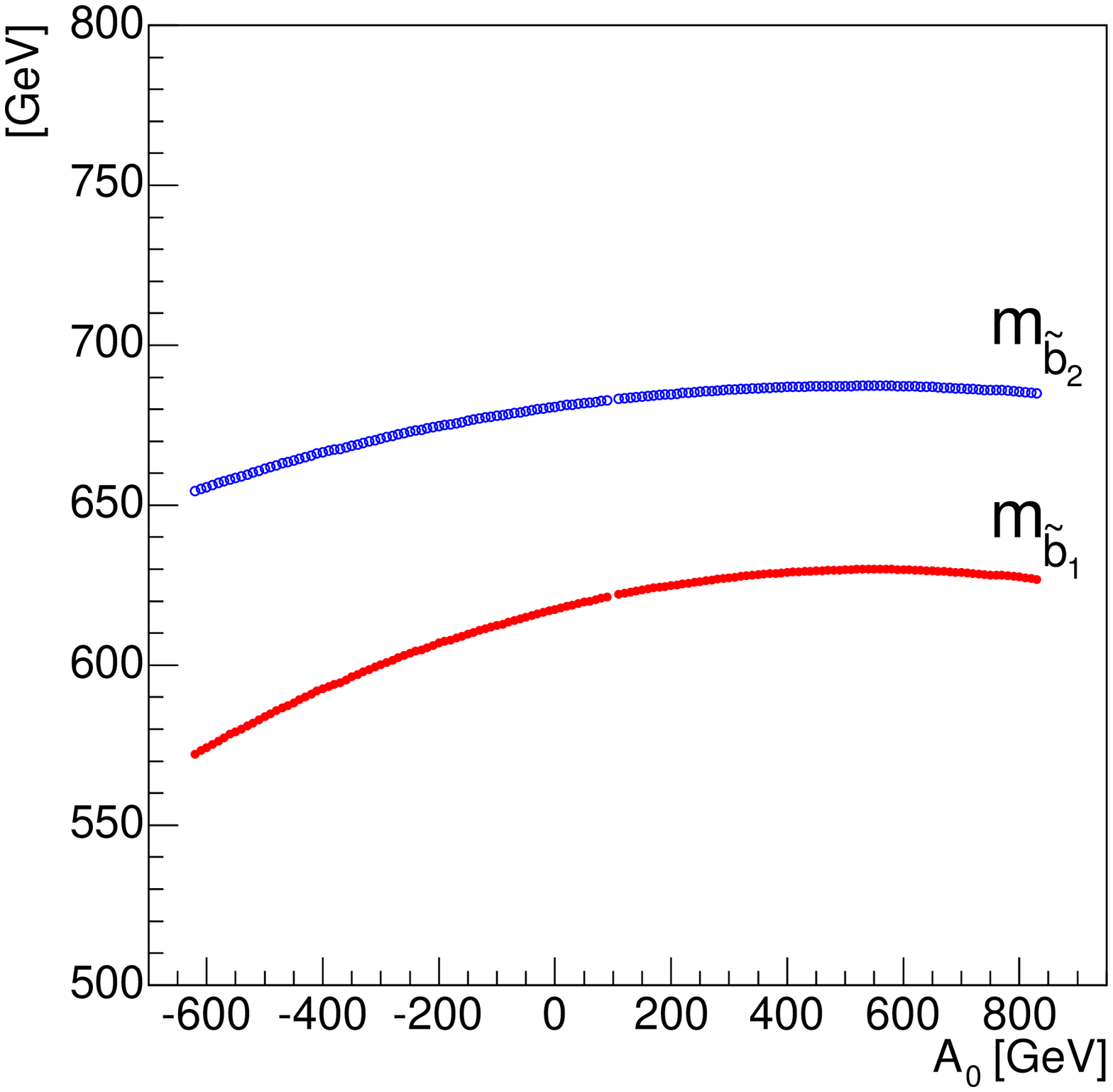}
\caption{\it Masses of the two sbottom squarks for the Snowmass point SPS4. The masses in the left/right plot are computed with SuSpect/ISAJET.}
\label{SPS4_mb}
\end{center}
\end{figure}

\begin{figure}[!h]
\vskip-.7cm
\begin{center}
\hspace*{-.3cm}\includegraphics[width=.50\textwidth]{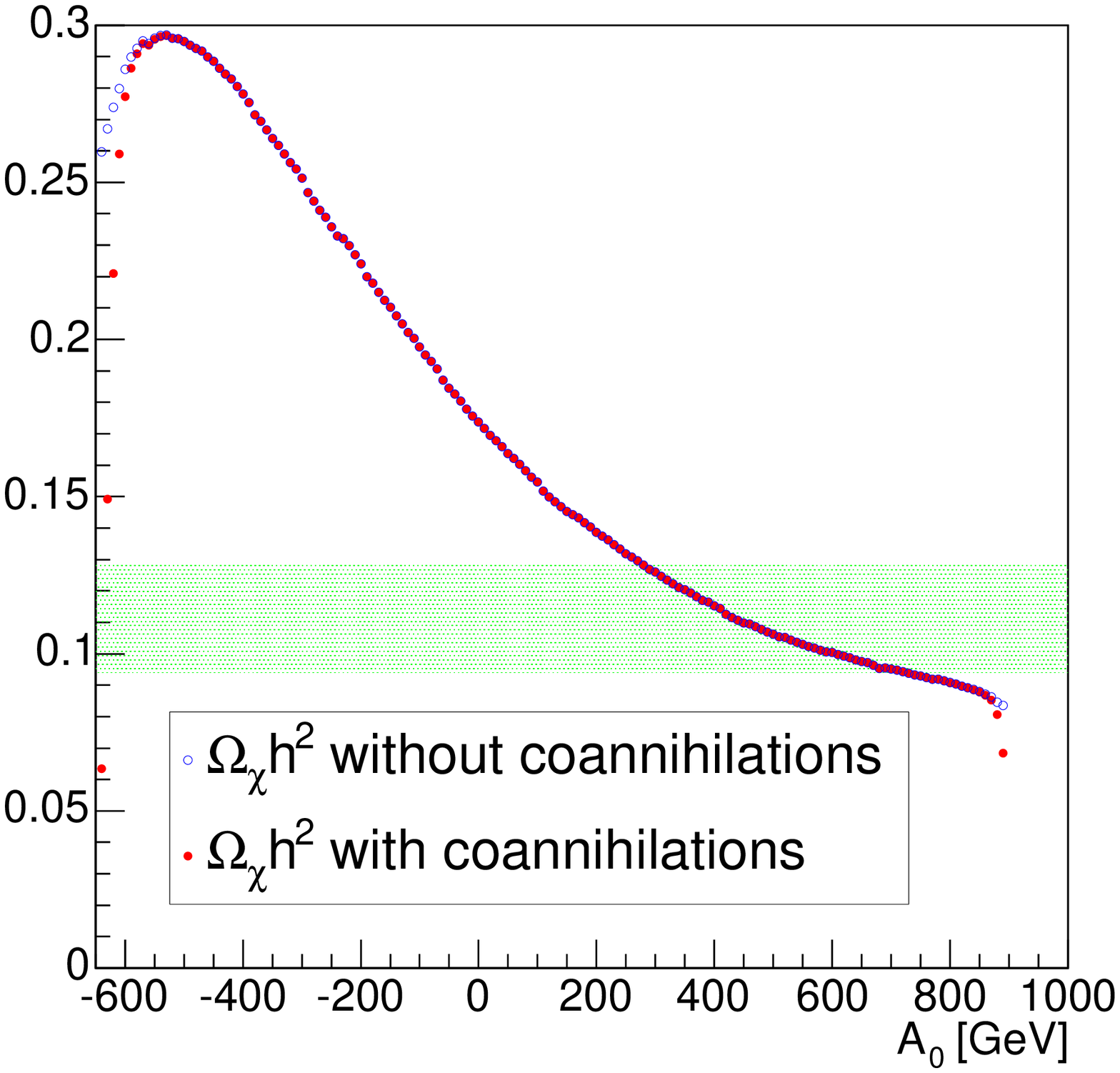}
\hspace*{-.3cm}\includegraphics[width=.50\textwidth]{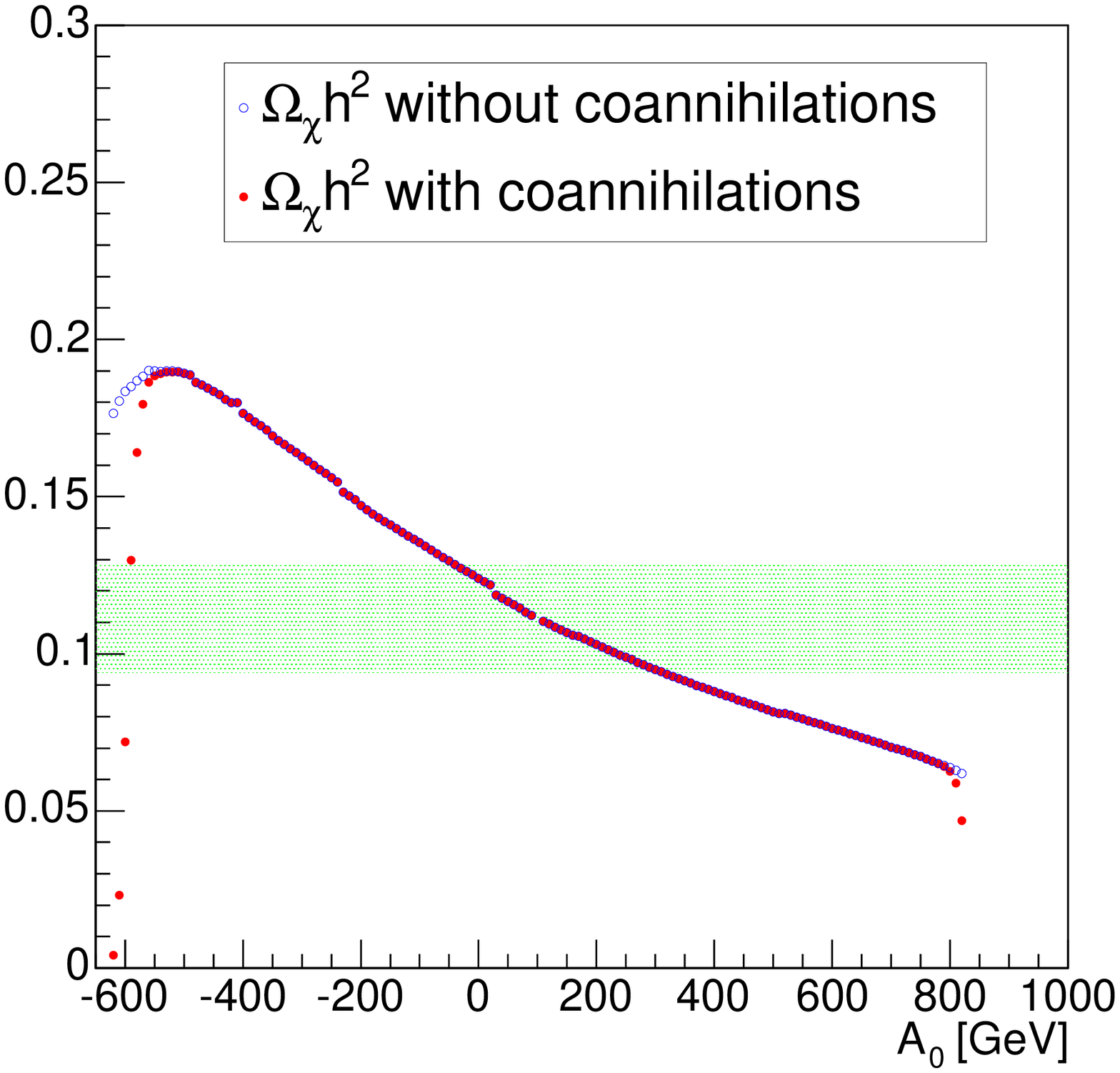}
\caption{\it Relic density as a function of $A_0$, obtained with the SuSpect (left) and ISAJET (right) programs for the Snowmass point SPS4. The blue circles correspond to a calculation without any coannihilation processes, while the red points include coannihilations. The green shaded region marks the allowed regions by the WMAP data.}
\label{SPS4_oh2}
\end{center}
\end{figure}

\noindent 
As the implementation of the RG evolution is different in the two programs, small differences in the SUSY spectra and couplings arise as shown in Figs.\ref{SPS4_mchi} and \ref{SPS4_mb}\footnote{Models with $A_0$ larger or smaller than displayed are excluded because of problems in the numerical application of the RG equations.}. The mass spectra provided by these two programs agree at a level of about $10\%$ for models with $m_0$ not much larger than $\m12$ and not too large values of $\tan\beta$. However, the calculated SUSY masses can differ by more than a factor of 2 for models in the focus point region or large $\tan \beta$ values\cite{ghodbane}. The neutralino masses calculated by the two programs agree quite well, e.g. for the Snowmass point SPS4 ($m_0$~=~400~GeV, $\m12$~=~300~GeV, tan$\beta$~=~50, $\mu >$~0): in Fig.\ref{SPS4_mchi} the masses of the four neutralinos are shown as a function of $A_0$. The LSP mass is independent of $A_0$ and the approximate mass degeneration as well as the $A_0$ dependence of the two heavy neutralinos is clearly visible.
\begin{figure}[!h]
\begin{center}
\hspace*{-.7cm}\includegraphics[width=1.\textwidth]{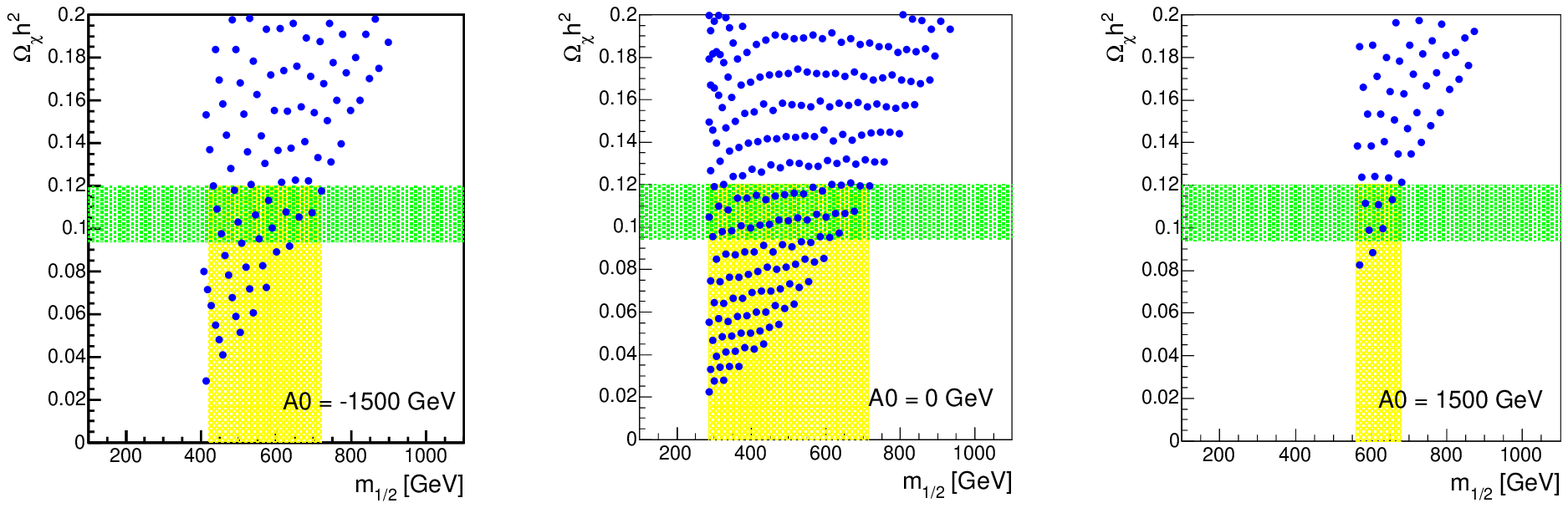}
\hspace*{-.7cm}\includegraphics[width=1.\textwidth]{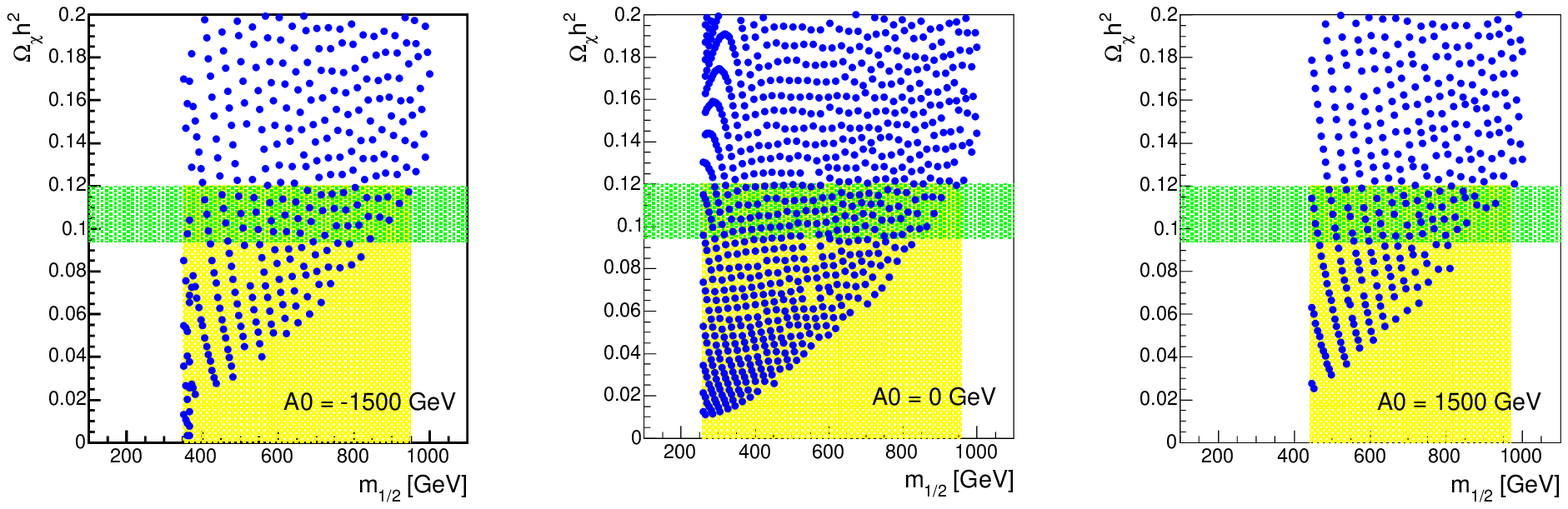}
\caption{\it Relic density as a function of $\m12$ obtained with the SuSpect (top row) and ISAJET (bottom row) programs for tan$\beta$~=~10, $\mu >$~0 and different values of $A_0$. The green shaded regions mark the WMAP allowed region. The yellow shaded regions show the corresponding $\m12$ ranges, which can differ significantly for the two MC programs.}
\label{oh2_comp}
\end{center}
\end{figure}
\noindent
However, for the same SPS benchmark point the masses of the sbottom squarks agree only on the $10\%$ level between the two programs, as indicated in Fig.\ref{SPS4_mb}. This difference directly affects the relic density calculations because of significant sbottom exchange contributions in the neutralino annihilation process. Due to the smaller sbottom mass $m_{\tilde{b}_1}$ calculated by SuSpect, the resulting relic density (Fig.\ref{SPS4_oh2}) is larger than the corresponding ISAJET value. Since the WMAP constraints on the relic density are quite strong, the number of models, which survive these constraints are different for SuSpect and ISAJET as demonstrated in Fig.\ref{oh2_comp}. Moreover, the regions in $\m12$, allowed by the WMAP data for fixed values of $A_0$, are different, too.


\section{Scan of the supersymmetric parameter space}
\label{scan}
Assuming CDM to consist exclusively of neutralinos, the cosmological bounds on the neutralino relic density $\Omega_{\chi} h^2$ imply strong constraints on the mSUGRA
parameter space. Since the aim of this analysis is to investigate the impact of $A_0$ on
the allowed regions in the mSUGRA parameter space, we varied $A_0$ in a first step, while keeping the other four parameters fixed.  For this study we
have chosen the Snowmass points \cite{snow} as benchmarks. In a second step we also varied
$m_0$ and $\m12$ for different but fixed values of $\tan \beta$. 
Recent data for $(g-2)_{\mu}$ favour a positive value of the Higgsino mass parameter \cite{g_2_mu}, therefore we focused our analysis on $\mu >$~0.
A few million mSUGRA models have been generated for the remaining four input parameters. We varied $m_0$ and $\m12$ between 0 and 2~TeV, $\tan \beta$ between 5 and 50 and $A_0$ within $\pm$4~TeV.  For the calculations the mass of the top quark has been set to 178~GeV.
\\
For all these models the relic density $\Omega_{\chi}$ has been computed and required to be within 0.094~$<\Omega_{\chi} h^2<$~0.129. In addition, the accelerator constraints adopted from the 2002 limits of the Particle Data Group have been taken into account \cite{dpg2002}. There are some updates of these constraints \cite{darksusy04}, including the rare decay $b \rightarrow s \gamma$. The accelerator constraints are applied to the mass spectra at the EW scale.

\begin{figure}[!h]
\begin{center}
\hspace*{-.8cm}\includegraphics[width=.54\textwidth]{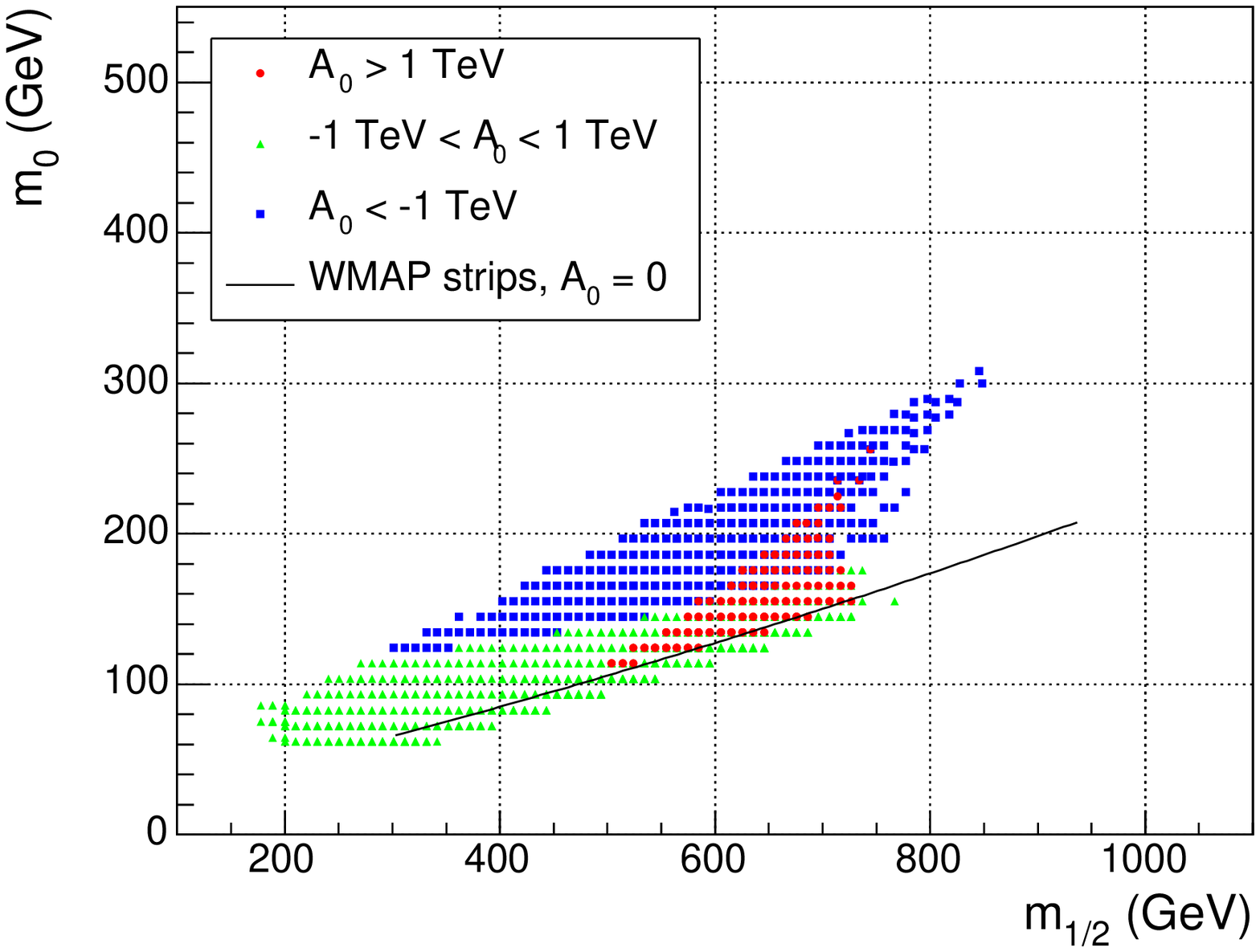}
\hspace*{-.7cm}\includegraphics[width=.54\textwidth]{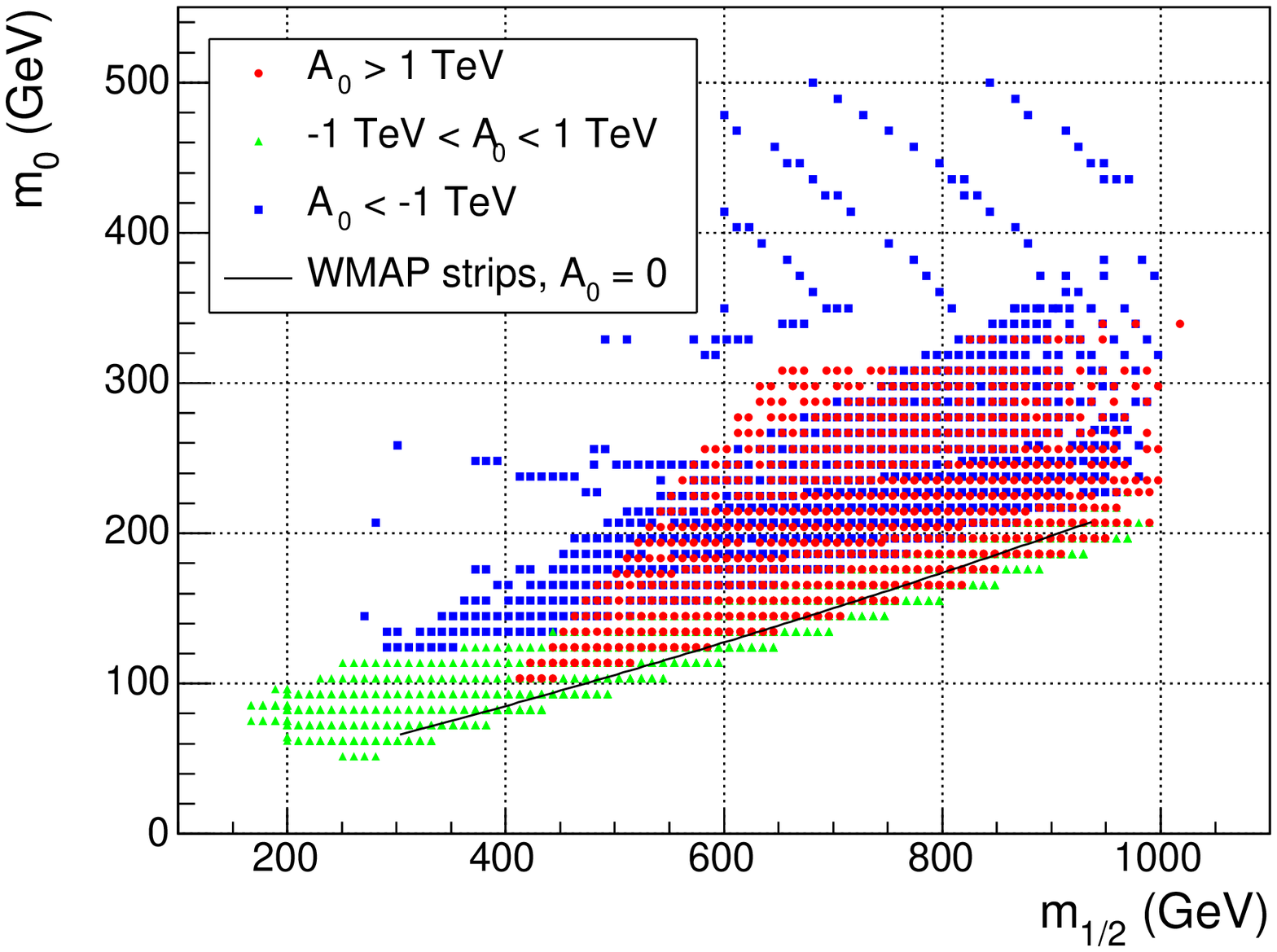}
\caption{\it Allowed models in the $m_0-\m12$ plane for tan$\beta$~=~10 and $\mu > 0$ obtained with SuSpect (left) / ISAJET (right). The black line corresponds to the WMAP strip for $A_0$~=~0~TeV from Ref.\cite{ellis_postW}.}
\label{C1_ranges}
\end{center}
\end{figure}
\noindent
Fig.\ref{C1_ranges} shows that the WMAP strip described in \cite{ellis_postW} broadens significantly, if $A_0$ is allowed to vary. The extension of the allowed regions originates mainly from larger values of $m_0$, which are allowed for non-vanishing $A_0$, while the minimal $m_0$ values are obtained for $A_0$~=~0~TeV. Large values of $\m12$ are excluded by the WMAP constraints on the relic density. Since slightly larger values of $\Omega_{\chi} h^2$ are derived by SuSpect than by ISAJET, more SuSpect models with larger $\m12$ are excluded. The parametrisation of the WMAP strips (black lines in Fig.\ref{C1_ranges}) have been determined with other programs \cite{ellis_postW}, thus introducing small differences with respect to our analysis.

\begin{figure}[!h]
\begin{center}
\hspace*{-.8cm}\includegraphics[width=.52\textwidth]{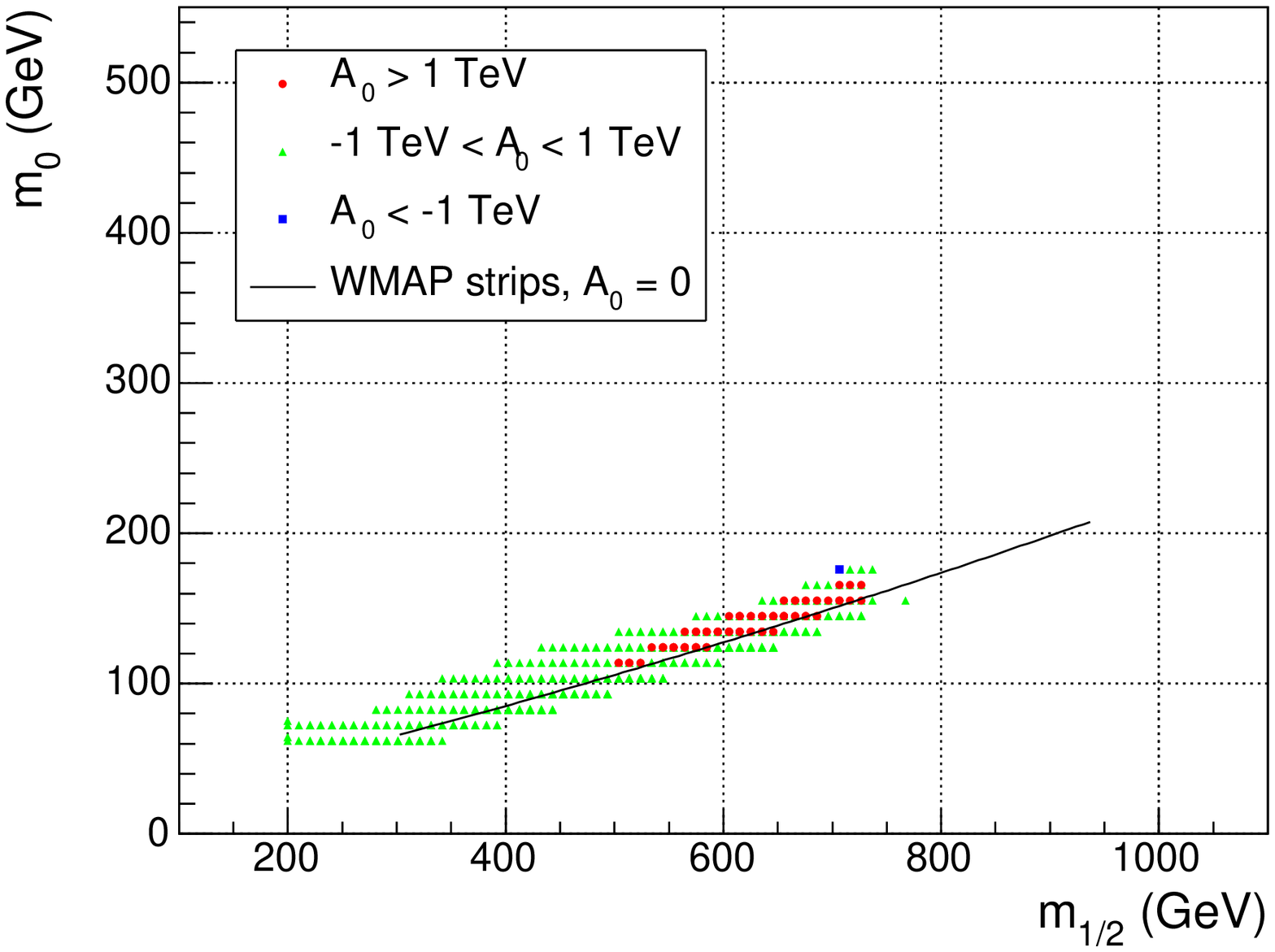}
\hspace*{-.0cm}\includegraphics[width=.52\textwidth]{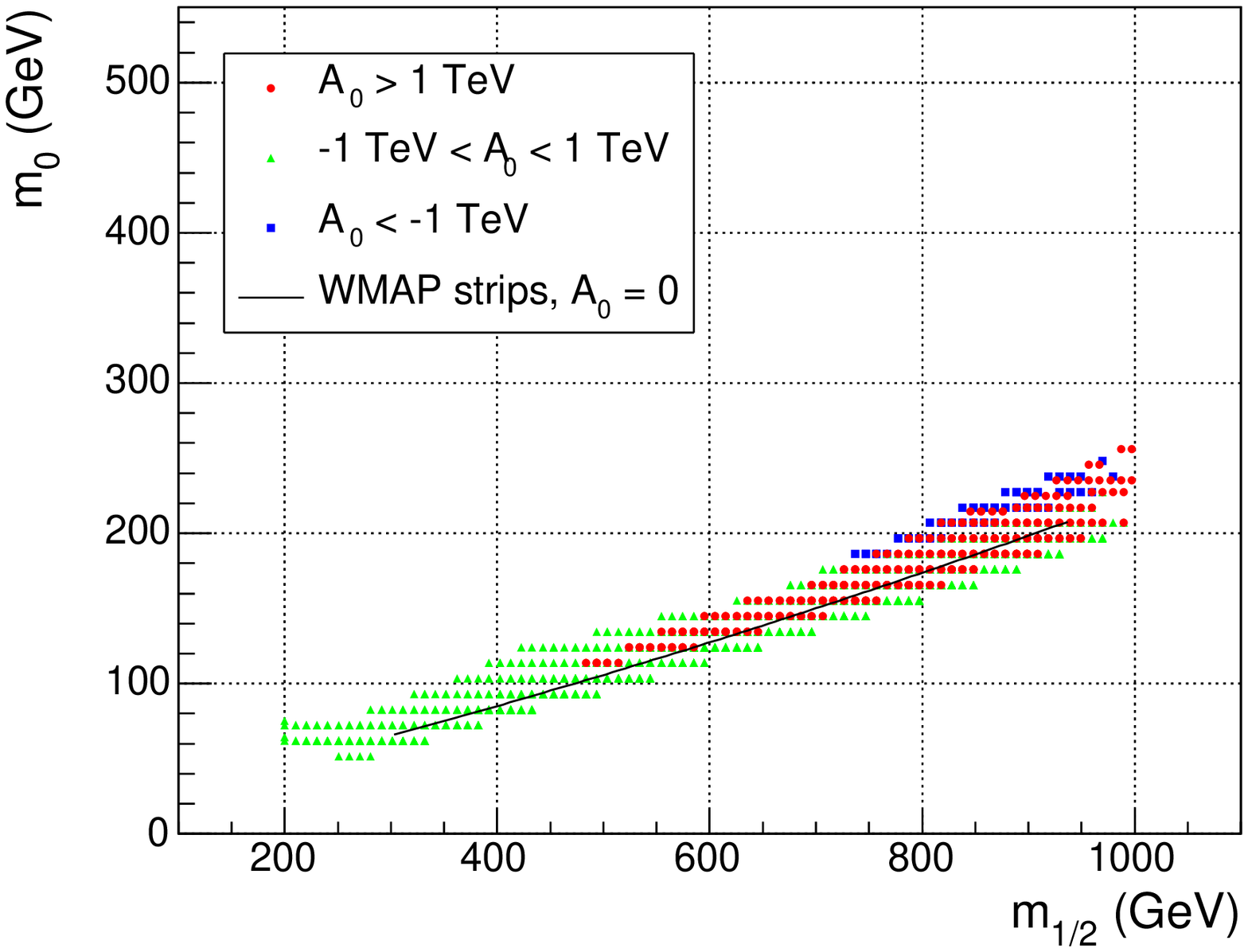}
\caption{\it  Allowed models in the $m_0-\m12$ plane for tan$\beta$~=~10 and $\mu > 0$ obtained with SuSpect (left) / ISAJET (right) after applying the cuts on $A_{t,b,\tau}$ to avoid colour and/or charge breaking. The black line corresponds to the WMAP strip for $A_0$~=~0~TeV from Ref.\cite{ellis_postW}. }
\label{C1_Acut}
\end{center}
\end{figure}

\noindent
In order to avoid colour
and/or charge breaking the trilinear scalar couplings at the EW scale $A_{t,b,\tau}$ have to be approximately constrained as \cite{ccb}:
\begin{eqnarray}
A_t^2 \le 3(m_{{H_u}}^2 + m_{\tilde{Q}_L}^2 + m_{\tilde{t}_R}^2),\nonumber\\
A_b^2 \le 3(m_{{H_d}}^2 + m_{\tilde{Q}_L}^2 + m_{\tilde{b}_R}^2),\\
A_{\tau}^2 \le 3(m_{{H_d}}^2 + m_{\tilde{L}_L}^2 + m_{\tilde{\tau}_R}^2).\nonumber
\label{cuts}\end{eqnarray}
\noindent
The consequences of applying these cuts is depicted in Fig.\ref{C1_Acut} for $\tan\beta$~=~10 and $\mu >$~0. About 35\% of the mSUGRA models, generated with SuSpect, satisfying the WMAP constraints are excluded, while for ISAJET the impact of these cuts is even stronger, i.e. about 45\% of the models are rejected (Fig.\ref{C1_Acut}). By far the biggest effect originates from the cut on $A_{\tau}$ due to the light stau masses $m_{\tilde{\tau}_{L,R}}$. Most of the models (more than 90\%) with $A_0<$~$-$1~TeV are excluded by the cuts. Models with $A_0$ values larger than 1~TeV are significantly affected, too (about 40\%).

\noindent
To avoid CCB these cuts are necessary but not sufficient, since the vacuum expectation values of the squarks, the sleptons and the corresponding Higgs boson were assumed to be equal, for simplicity. Moreover these bounds (Eq.\ref{cuts}) were derived from the tree level scalar potential, while radiative corrections are expected to modify them.\\
The scalar potential may contain global CCB minima in addition to the local EW breaking minima. As no CCB has been observed, the Universe in its present state may be trapped in a local EW breaking minimum. Since this metastable state may have a lifetime longer than the age of the Universe due to the small tunnelling probability into the global minimum \cite{riotto}, CCB cannot be excluded. Thus, we did not apply these cuts in this analysis.

\begin{figure}[!h]
\begin{center}
\hspace*{-.7cm}\includegraphics[width=.54\textwidth]{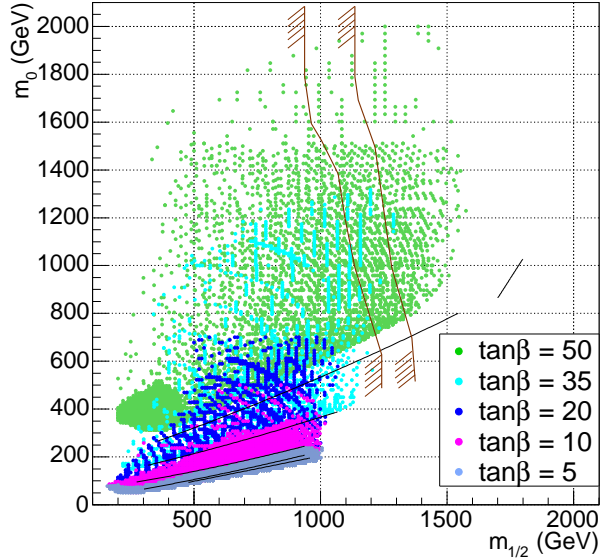}
\caption{\it  Allowed models in the $m_0-\m12$ plane for $m_0~,\m12\le$~2~TeV, tan$\beta$ between 5 and 50, $\mu~>~0$ and $A_0$ within $\pm$4~TeV from ISAJET. The black lines correspond to the WMAP strips as parametrised in \cite{ellis_postW}, with tan$\beta$~=~5, 10, 20, 35 and 50, where tan$\beta$~=~5 belongs to the lowest black line and tan$\beta$~=~50 to the highest one. The brown lines indicate the LHC discovery reach for an integrated luminosity of 100~fb$^{-1}$ and 300~fb$^{-1}$ \cite{CMS}, respectively.}
\label{C_all}
\end{center}
\end{figure}
\noindent
In Fig.\ref{C_all} all allowed models for different values of the trilinear scalar coupling $A_0$ and $\tan\beta$ are shown in the $m_0 - \m12$ plane assuming the Higgsino mass parameter $\mu$ to be positive. This plot corresponds to Fig.\ref{C1_ranges} but with $\tan\beta$~=~5, 10, 20, 35 and 50 superimposed. In contrast to Fig.\ref{Wstrips}, where only a few lines survived the WMAP constraints, extended regions in the mSUGRA parameter space are allowed, if $A_0$ is varied. For $\tan\beta$~=~10 mSUGRA models with $m_0 \sim$~350~GeV for largely negative $A_0$ are within the WMAP constraints, while for $A_0$~=~0~TeV the upper bound was about  $m_0 \sim$~200~GeV. The analogous behaviour can be observed for larger $\tan\beta$ values, where mSUGRA models with $m_0$ around 2~TeV are allowed for large $|A_0|$.
\begin{figure}[!h]
\begin{center}
\hspace*{-.8cm}\includegraphics[width=.52\textwidth]{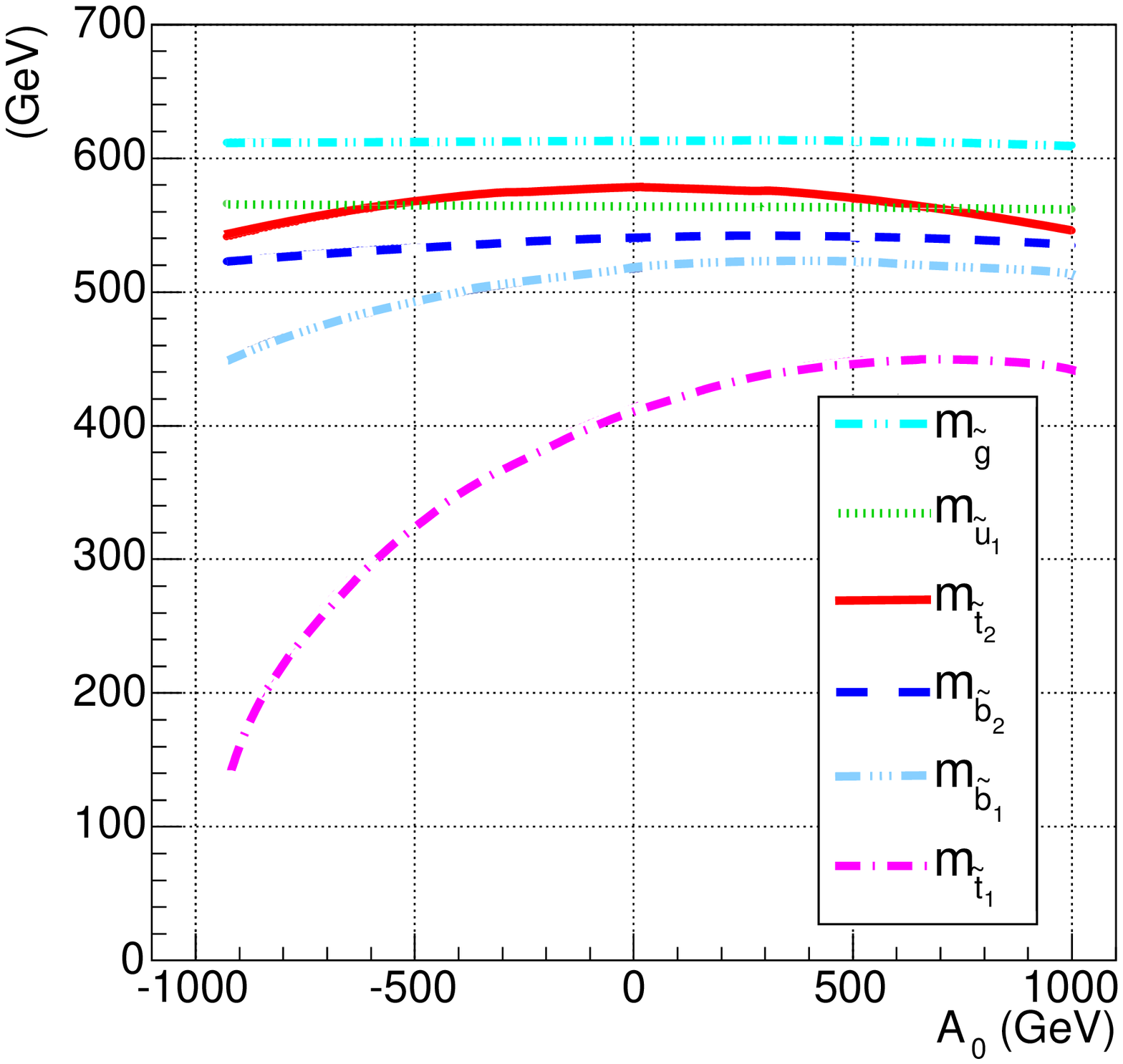}
\hspace*{-.0cm}\includegraphics[width=.52\textwidth]{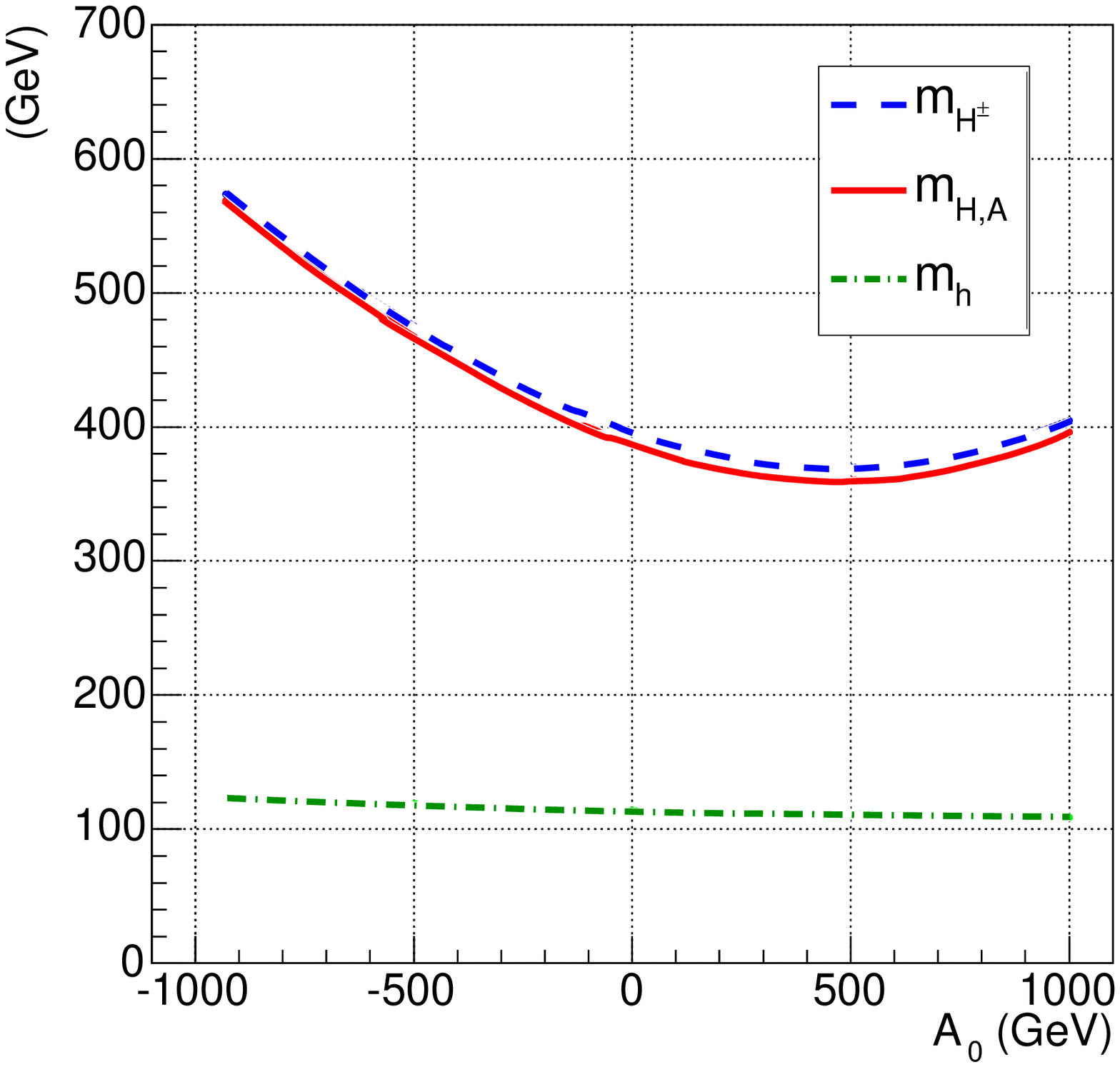}
\caption{\it Gluino, squark and Higgs masses for SPS1A ($m_0$~=~100~GeV, $\m12$~=~250~GeV, tan$\beta$~=~10, $\mu >$~0) as a function of $A_0$ obtained with ISAJET. }
\label{squark}
\end{center}
\end{figure}
\noindent
The masses of the gluinos, of the first two squark generations and of the light scalar Higgs boson are nearly independent of $A_0$. Since the squarks of the first two generations are almost degenerated in mass, only $m_{\tilde{u}_{1}}$ is shown in Fig.\ref{squark} as representative examples. However, the third generation squark masses $m_{\tilde{b}_{1,2}}$, $m_{\tilde{t}_{1,2}}$ as well as the heavier Higgs masses $m_{H,A,H^\pm}$ depend on the trilinear coupling $A_0$ in general. The future Large Hadron Collider (LHC), with an anticipated luminosity of 100~fb$^{-1}$, will cover the $\m12$ domain up to about $\mathcal{O}$(TeV), almost independent of $m_0$ \cite{CMS} (Fig.\ref{C_all}). Thus, most of the models generated in this analysis are within the reach of the LHC.


\section{Parametrisation}
\label{results}

\begin{figure}[!h]
\begin{center}
\hspace*{-.8cm}\includegraphics[width=.53\textwidth]{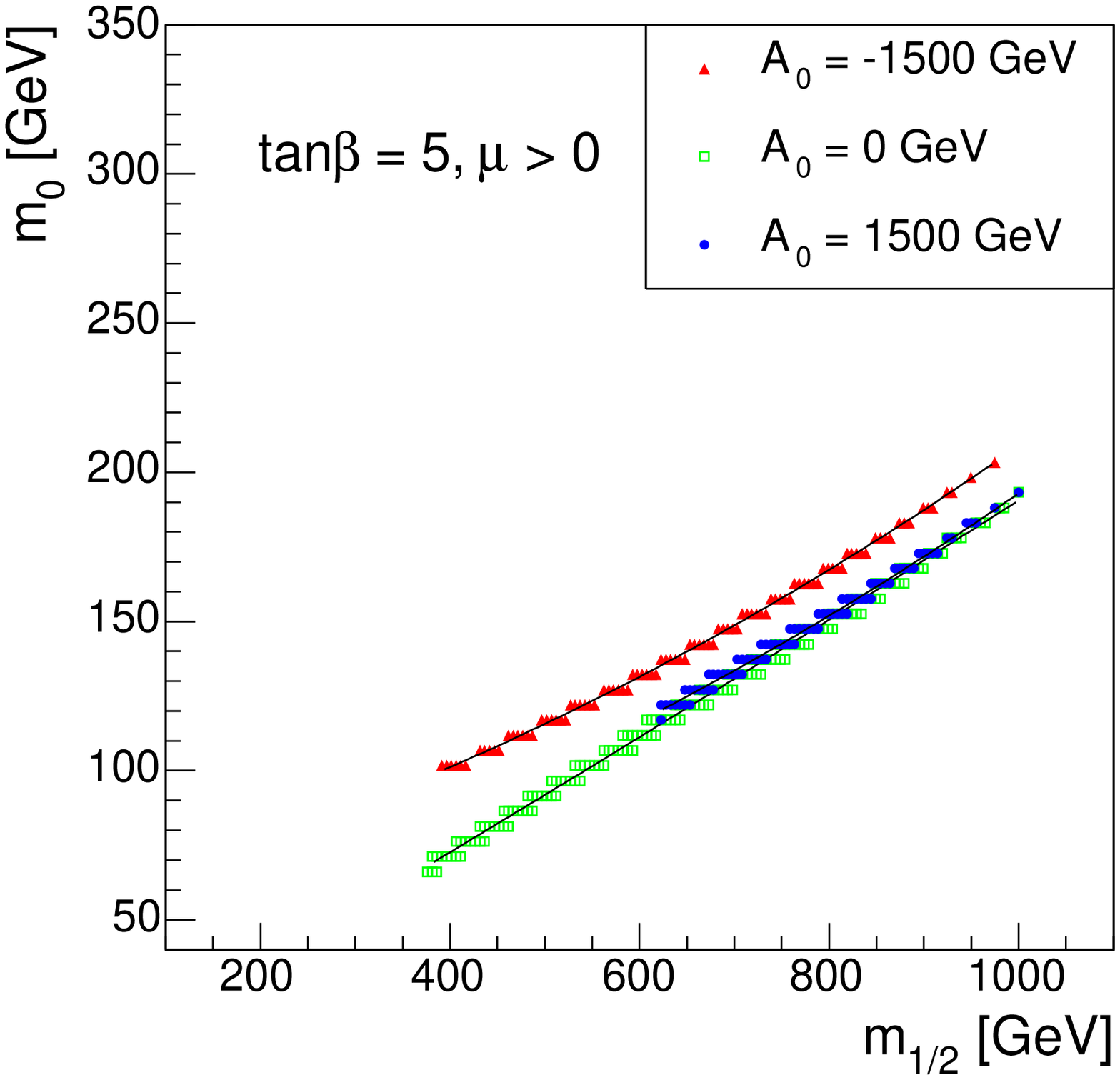}
\hspace*{-.6cm}\includegraphics[width=.53\textwidth]{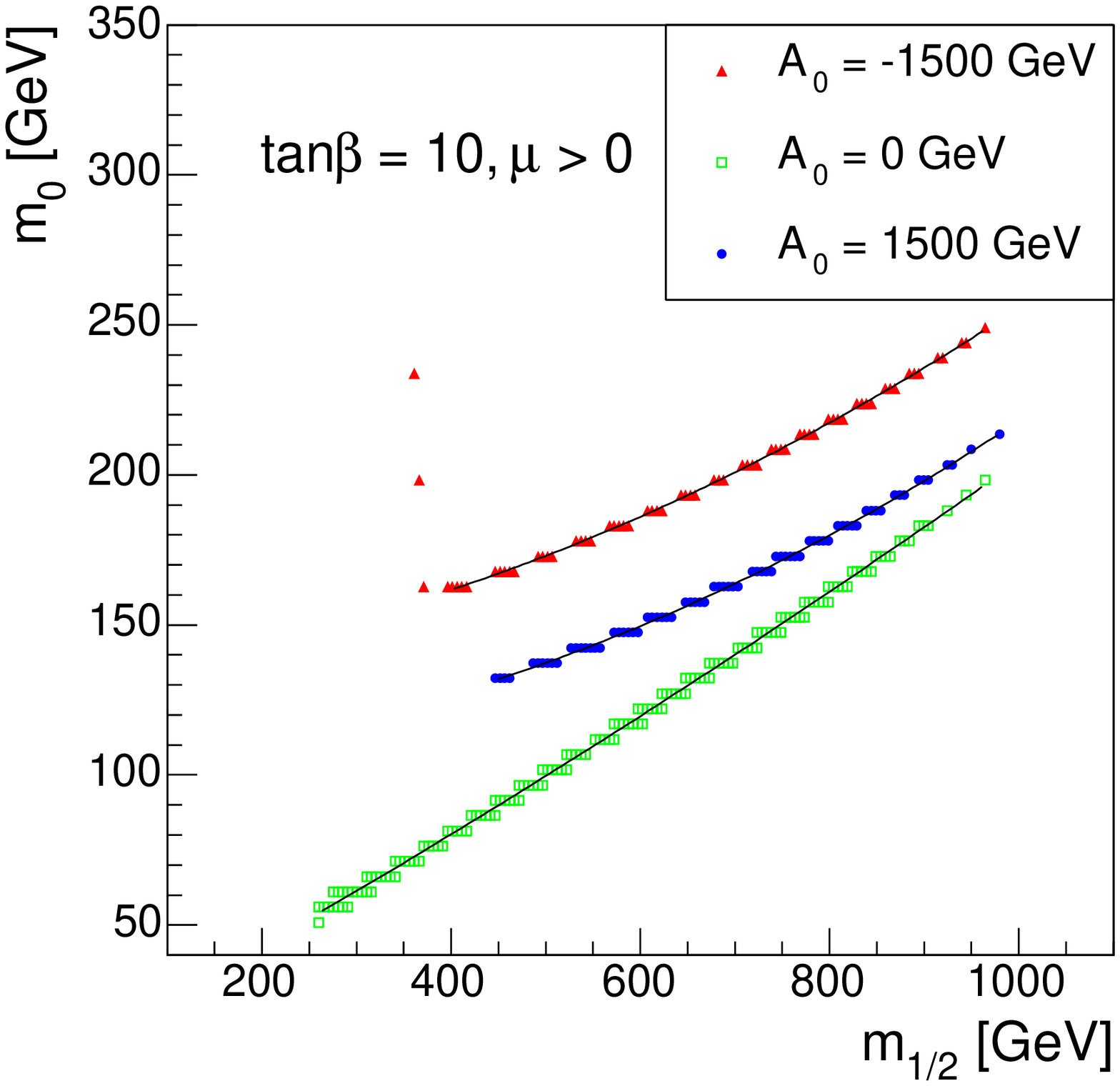}
\caption{\it Allowed regions in the $m_0-\m12$ plane, on the left for tan$\beta$~=~5 and on the right for tan$\beta$~=~10. In both plots $\mu$ is positive. The black lines are the fits given in Tabs.\ref{tab:tanbeta5} and \ref{tab:tanbeta10}.}
\label{parametrisation5_10}
\end{center}
\end{figure}

\begin{figure}[!h]
\begin{center}
\hspace*{-.8cm}\includegraphics[width=.53\textwidth]{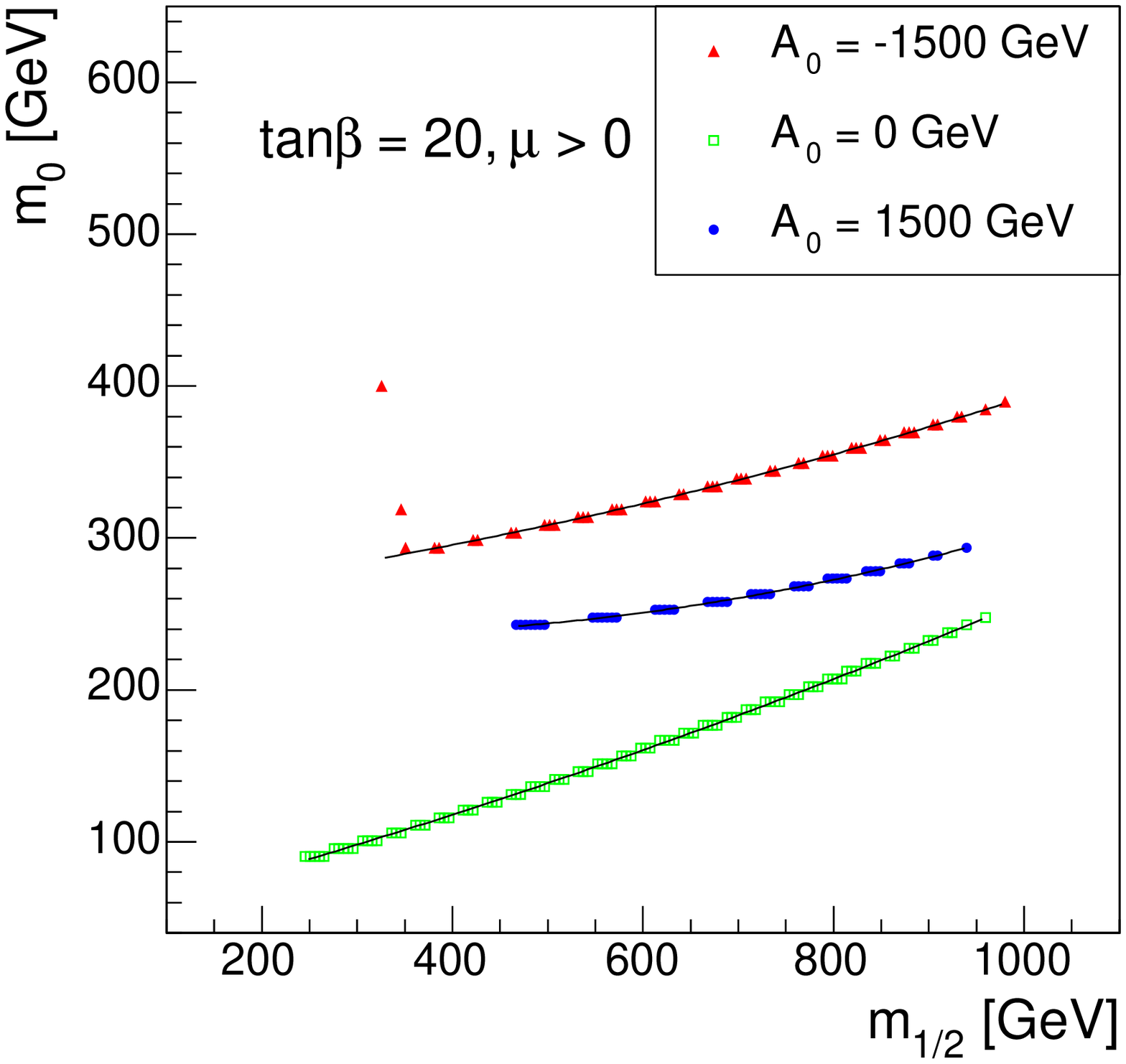}
\hspace*{-.6cm}\includegraphics[width=.53\textwidth]{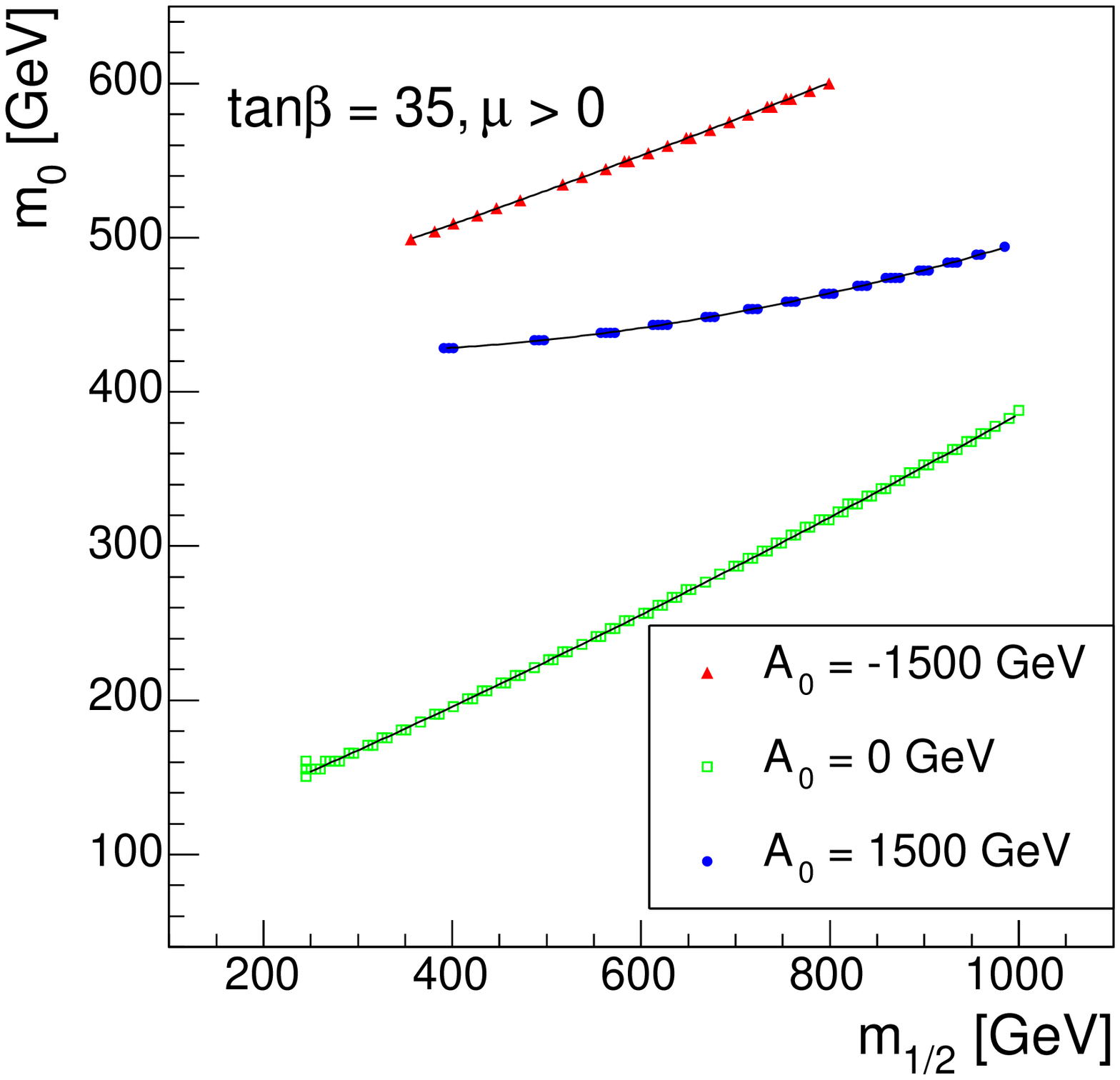}
\caption{\it Allowed regions in the $m_0-\m12$ plane for tan$\beta$~=~20 (left) and for tan$\beta$~=~35 (right). In both plots $\mu$ is positive. The black lines are the fits given in  Tabs.\ref{tab:tanbeta20} and \ref{tab:tanbeta35}.}
\label{parametrisation20_35}
\end{center}
\end{figure}

\begin{figure}[!h]
\begin{center}
\includegraphics[width=.53\textwidth]{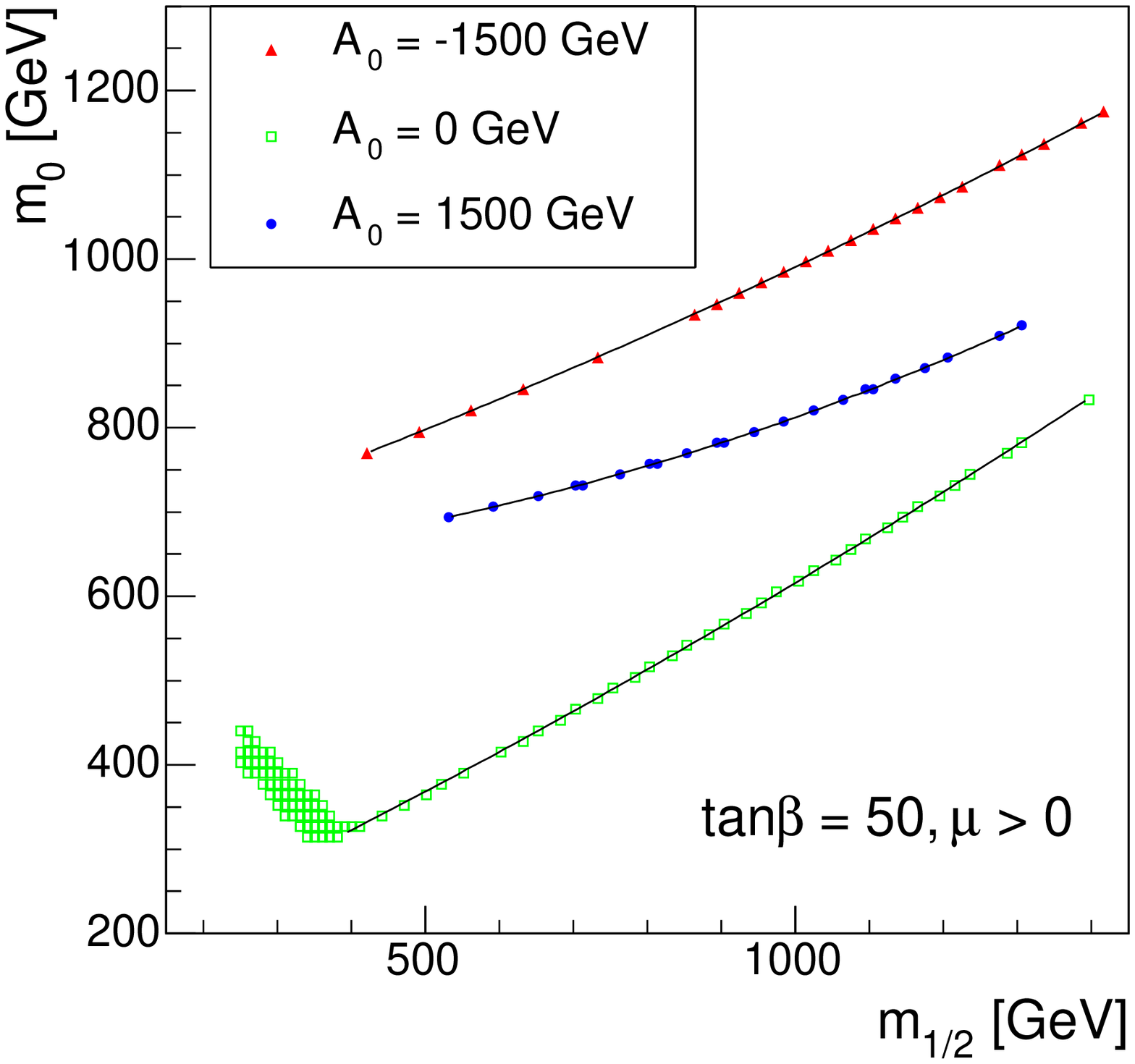}
\caption{\it Allowed regions in the $m_0-\m12$ plane for tan$\beta$~=~50 and $\mu > 0$. The black lines are the fits given in Tab.\ref{tab:tanbeta50}.}
\label{parametrisation50}
\end{center}
\end{figure}

As can be inferred from Figs.\ref{parametrisation5_10},  \ref{parametrisation20_35} and \ref{parametrisation50}  the allowed regions in the $m_0-\m12$ plane correspond to lines if the trilinear coupling $A_0$ is kept at fixed values. Because of their smooth narrow shape, they can be fitted by a polynomial of 2nd order: 
 \begin{eqnarray*}
 m_0 = a + b\cdot\m12+ c\cdot\m12^2.
 \end{eqnarray*}
  The coefficients of the quadratic terms are small, but they should not be neglected as $c\cdot \m12^2$ can be of the same magnitude as $b \cdot \m12$ in the allowed $\m12$ domain: the surviving mSUGRA models do not lie on a straight line but on a curve (Figs.\ref{parametrisation5_10} to\ref{parametrisation50}).
 The parameters we obtained by using the MINUIT \cite{minuit} routines are given in Tabs.\ref{tab:tanbeta5} to \ref{tab:tanbeta50} for different values of $\tan\beta$ and $A_0$. The numbers have been calculated from the ISAJET output.

\begin{table}[!h]
\begin{tabular}{r c c c c c}
        \multicolumn{1}{c}{$A_0$} & a & b & c & $\m12$ domain & $m_0$ domain\\
        \hline 
        -2000 & 86 $\pm$ 4 & \hspace*{-.15cm}-0.04 $\pm$ 0.01 & 0.000089 $\pm$ 0.000008 & 502 - 955~GeV & 132 - 208~GeV\\
        -1500 & 57 $\pm$ 2&  0.081 $\pm$ 0.006 & 0.000007 $\pm$ 0.000004 & 391 - 975~GeV & 102 - 203~GeV\\
        -1000 & 35 $\pm$ 1 & 0.110 $\pm$ 0.003 & 0.000056 $\pm$ 0.000002 &  265 - 985~GeV & 71 - 198~GeV\\
         -500 & \hspace*{.15cm}9.4 $\pm$ 0.7 & 0.161 $\pm$ 0.002 & 0.000025 $\pm$ 0.000002 & 260 - 985~GeV & 51 - 193~GeV\\
            0 & -2 $\pm$ 1 & 0.184 $\pm$ 0.004 & 0.000009 $\pm$ 0.000003 & \hspace*{.15cm}380 - 1000~GeV & 66 - 193~GeV\\
          500 & 0 $\pm$ 2 & 0.177 $\pm$ 0.007 & 0.000013 $\pm$ 0.000005 & 482 - 990~GeV & 86 - 188~GeV\\
         1000 & 15 $\pm$ 4 & 0.14 $\pm$ 0.01 & 0.000035 $\pm$ 0.000008 & 562 - 990~GeV & 102 - 188~GeV\\
         1500 & 44 $\pm$ 8 & 0.08 $\pm$ 0.02 & 0.00007 $\pm$ 0.00001 & \hspace*{.15cm}623 - 1000~GeV & 117 - 193~GeV\\
         2000 & \hspace*{.15cm}76 $\pm$ 14 & 0.02 $\pm$ 0.03 & 0.00010 $\pm$ 0.00002  & \hspace*{.15cm}683 - 1000~GeV & 137 - 198~GeV\\
         2500 & \hspace*{.15cm}85 $\pm$ 26 & 0.03 $\pm$ 0.06 & 0.00009 $\pm$ 0.00004  & 733 - 990~GeV & 158 - 203~GeV\\
\end{tabular}
\caption[]{\label{tab:tanbeta5}  \it Coefficients $a, b$ and $c$ of the parametrisation for tan$\beta$~=~5 for different discrete values of $A_0$ between $-$2~TeV and $+$2.5~TeV. For larger or smaller $A_0$ values too few mSUGRA models survive the WMAP constraints to allow a reasonable parametrisation. The last two columns contain the domains for $m_{1/2}$ and $m_0$, respectively.}
\end{table}

\begin{table}[!h]
\begin{tabular}{r c c c c c}
        \multicolumn{1}{c}{$A_0$} & a & b & c & $\m12$ domain & $m_{0}$ domain\\
        \hline 
        -2000 & 189 $\pm$ 4 &  \hspace*{.1cm}-0.02 $\pm$ 0.01  & 0.000113  $\pm$ 0.000009  &  485 - 965~GeV & 208 - 285~GeV\\
        -1500 & 134 $\pm$ 2\hspace*{.25cm}  &  \hspace*{.15cm}0.034 $\pm$ 0.007  & 0.000088  $\pm$ 0.000005  & 400 - 965~GeV & 163 - 249~GeV\\
        -1000 &  96 $\pm$ 1  &  \hspace*{.35cm}0.045 $\pm$ 0.004 & 0.000094 $\pm$ 0.000003 & 260 - 954~GeV & 117 - 224~GeV\\
         -500 &  47 $\pm$ 1  & 0.104 $\pm$ 0.003\hspace*{-.2cm} & 0.000066 $\pm$ 0.000003 & 260 - 939~GeV & 76 - 203~GeV\\
            0 &  \hspace*{.05cm} 8 $\pm$ 1  & 0.171 $\pm$ 0.003\hspace*{-.2cm} & 0.000026 $\pm$ 0.000002 & 260 - 964~GeV & 51 - 198~GeV\\
          500 &  11 $\pm$ 1  & 0.157 $\pm$ 0.004\hspace*{-.2cm} & 0.000033 $\pm$ 0.000003 & 340 - 980~GeV & 66 - 198~GeV\\
         1000 &  53 $\pm$ 2  & 0.081 $\pm$ 0.006\hspace*{-.2cm} & 0.000072 $\pm$ 0.000004 & 390 - 984~GeV & 97 - 203~GeV\\
         1500 & 105 $\pm$ 3\hspace*{.25cm}  & 0.016 $\pm$ 0.009\hspace*{-.2cm} & 0.000097 $\pm$ 0.000006 & 447 - 980~GeV & 132 - 214~GeV\\
         2000 & 154 $\pm$ 4\hspace*{.25cm}  & -0.02 $\pm$ 0.01  & 0.000103 $\pm$ 0.000008 & 487 - 960~GeV & 168 - 229~GeV\\
         2500 & 200 $\pm$ 6\hspace*{.25cm}  & -0.05 $\pm$ 0.02  & 0.00010  $\pm$ 0.00001  & 562 - 1000~GeV\hspace*{-.2cm} & 208 - 259~GeV\\
\end{tabular}
\vskip.4cm
\caption[]{\label{tab:tanbeta10} \it Same as in Tab.\ref{tab:tanbeta5}, but for tan$\beta$~=~10.}
\end{table}

\noindent
The gaps on the fitted lines for $\tan \beta$~=~50 originate from the chosen step size for $m_0$ (Fig.\ref{oh2_form}). Some plots exhibit extended areas for small $m_0$ and small $\m12$ (for example Fig.\ref{parametrisation50} for $A_0$~=~0~TeV). These regions are excluded in the fits, since a proper parametrisation would be much more involved.

\begin{table}[!h]
\begin{tabular}{r c c c c c c}
       \multicolumn{1}{c}{$A_0$} & a & b & c & $\m12$ domain & $m_0$ domain\\
        \hline 
        -1500 & 257 $\pm$ 2  &  0.071  $\pm$ 0.008  & 0.000066 $\pm$ 0.000006 & 326 - 980~GeV & 293 - 400~GeV\\
        -1000 & 183 $\pm$ 2  &  0.080 $\pm$ 0.005 & 0.000074 $\pm$ 0.000004 & 286 - 965~GeV & 212 - 329~GeV\\
         -500 & 108 $\pm$ 1  &  0.106 $\pm$ 0.003 & 0.000076 $\pm$ 0.000003 & 220 - 975~GeV & 136 - 283~GeV\\
            0 &  \hspace*{.15cm}46 $\pm$ 1  &  0.160 $\pm$ 0.003 & 0.000053 $\pm$ 0.000003 & 245 - 960~GeV & 91 - 248~GeV\\
          500 &  \hspace*{.15cm}68 $\pm$ 1  &  0.100 $\pm$ 0.005 & 0.000083 $\pm$ 0.000004 & 326 - 970~GeV & 111 - 243~GeV\\
         1000 & 153 $\pm$ 3  & \hspace*{-.15cm}-0.005 $\pm$ 0.008 & 0.000119 $\pm$ 0.000006 & 401 - 960~GeV & 172 - 258~GeV\\
         1500 & 246 $\pm$ 4  &  -0.07  $\pm$ 0.01  & 0.000125 $\pm$ 0.000009 & 467 - 940~GeV & 243 - 294~GeV\\
         2000 & 327 $\pm$ 3  & \hspace*{-.15cm}-0.079 $\pm$ 0.009 & 0.000107 $\pm$ 0.000007 & 447 - 985~GeV & 314 - 354~GeV\\
         2500 & \hspace*{.15cm}410 $\pm$ 18 & \hspace*{-.15cm}-0.09  $\pm$ 0.05  & 0.00010  $\pm$ 0.00004  & 557 - 819~GeV & 390 - 400~GeV\\
\end{tabular}
\vskip.4cm
\caption[]{\label{tab:tanbeta20}  \it Same as in Tab.\ref{tab:tanbeta5}, but for tan$\beta$~=~20.}
\end{table}

\begin{table}[!h]
\begin{tabular}{r c c c c c c}
        \multicolumn{1}{c}{$A_0$} & a & b & c & $\m12$ domain & $m_0$ domain\\
        \hline 
         -1500 & 427 $\pm$ 5 &  0.19  $\pm$ 0.02  & 0.00004  $\pm$ 0.00001  & 356 - 799~GeV & 499 - 600~GeV\\
         -1000 & 301 $\pm$ 2 &  0.190 $\pm$ 0.005 & 0.000051 $\pm$ 0.000004 & 281 - 995~GeV & 358 - 540~GeV\\
          -500 & 176 $\pm$ 1 &  0.219 $\pm$ 0.004 & 0.000054 $\pm$ 0.000003 & \hspace*{.15cm}245 - 1000~GeV & 231 - 448~GeV\\
             0 & \hspace*{.1cm} 88 $\pm$ 1 &  0.251 $\pm$ 0.003 & 0.000047 $\pm$ 0.000002 & \hspace*{.15cm}245 - 1000~GeV & 151 - 388~GeV\\
           500 & 138 $\pm$ 1 &  0.139 $\pm$ 0.004 & 0.000097 $\pm$ 0.000003 & \hspace*{.15cm}311 - 1000~GeV & 191 - 373~GeV\\
          1000 & 282 $\pm$ 2 &  0.001 $\pm$ 0.007 & 0.000131 $\pm$ 0.000005 & 376 - 995~GeV & 302 - 413~GeV\\
          1500 & 432 $\pm$ 3 & \hspace*{-.1cm}-0.058 $\pm$ 0.009 & 0.000122 $\pm$ 0.000006 & 391 - 985~GeV & 428 - 494~GeV\\
          2000 & 577 $\pm$ 5 & \hspace*{-.1cm}-0.08  $\pm$ 0.01  & 0.000103  $\pm$ 0.000010  & 517 - 980~GeV & 565 - 600~GeV\\
\end{tabular}
\vskip.4cm
\caption[]{\label{tab:tanbeta35}  \it Same as in Tab.\ref{tab:tanbeta5}, but for tan$\beta$~=~35.}
\end{table}

\noindent
The relic density develops a strong dependence on the common gaugino mass $\m12$ for small values near the lower bound coming from the accelerator constraints. Consequently, unexpectedly large $m_0$ values (for small $\m12$) can lead to $\Omega_{\chi}h^2$ values within the WMAP range. This behaviour manifests itself e.g. in the isolated points in Fig.\ref{parametrisation5_10} ($\tan\beta$~=~10) and in Fig.\ref{parametrisation20_35} ($\tan\beta$~=~20), for $A_0$~=~$-$1500~GeV.

\begin{table}[!h]
\begin{tabular}{r c c c c c c}
         \multicolumn{1}{c}{$A_0$} & a & b & c & $\m12$ domain &$m_0$ domain\\
        \hline 
        -2000 &  \hspace*{.15cm}830 $\pm$ 6 &  0.30  $\pm$ 0.02  & 0.00004  $\pm$ 0.00001  & \hspace*{.15cm}311 - 1085~GeV & 922 - 1200~GeV\\
        -1500 &  635 $\pm$ 3  &  0.294 $\pm$ 0.006 & 0.000061 $\pm$ 0.000003 & 421 - 1416~GeV & 770 - 1175~GeV\\
        -1000 &  425 $\pm$ 5  &  0.35  $\pm$ 0.01  & 0.000055 $\pm$ 0.000005 & 472 - 1427~GeV & 605 - 1035~GeV\\
         -500 &  249 $\pm$ 1  &  0.395 $\pm$ 0.003 & 0.000054 $\pm$ 0.000002 & 290 - 1427~GeV & 352 - 922~GeV\\
            0 &  152 $\pm$ 1  &  0.401 $\pm$ 0.004 & 0.000063 $\pm$ 0.000002 & 390 - 1397~GeV & 314 - 833~GeV\\
          500 &  236 $\pm$ 2  &  0.250 $\pm$ 0.005 & 0.000122 $\pm$ 0.000003 & 430 - 1286~GeV & 365 - 757~GeV\\
         1000 &  429 $\pm$ 1  &  0.100 $\pm$ 0.004 & 0.000153 $\pm$ 0.000002 & 341 - 1286~GeV & 478 - 808~GeV\\
         1500 &  630 $\pm$ 5  &  0.05  $\pm$ 0.01  & 0.000132 $\pm$ 0.000005 & 532 - 1306~GeV & 694 - 922~GeV\\
         2000 &  852 $\pm$ 4  &  0.000 $\pm$ 0.009 & 0.000122 $\pm$ 0.000005 & 411 - 1306~GeV & 871 - 1061~GeV\\
         2500 & 1073 $\pm$ 39 & \hspace*{-.15cm}-0.03  $\pm$ 0.08  & 0.00011  $\pm$ 0.00004  & 653 - 1155~GeV & 1099 - 1187~GeV\\
\end{tabular}
\vskip.4cm
\caption[]{\label{tab:tanbeta50}  \it Same as in Tab.\ref{tab:tanbeta5}, but for tan$\beta$~=~50.}
\end{table}

\noindent
The separation between the lines for different $A_0$ becomes larger with increasing $\tan\beta$ and increasing $|A_0|$ (Figs.\ref{parametrisation5_10} to \ref{parametrisation50}). For $\tan\beta$~=~5 the corresponding splitting in $m_0$ is smaller than 90~GeV, while for $\tan\beta$ =~50 it can reach about 700~GeV. For negative $A_0$ the shift in $m_0$ is larger than for positive values, but always to higher $m_0$ values so that the minimal $m_0$ values are obtained for vanishing $A_0$. For small $\tan\beta$ the dominant contribution to the annihilation cross section may come from the t-channel stau exchange (especially in the coannihilation region $m_0~\ll\m12$). The running of the stau masses depends quadratically on $A_0$. For large values of $\tan\beta$ the rapid annihilation via a s-channel Higgs boson exchange can be dominant. The Higgs masses depend linearly as well as a quadratically on $A_0$ \cite{ellisA0}.
These dependencies of the sparticles and  Higgs bosons masses on $A_0$ (Fig.\ref{squark}) affect the cross sections and therefore also the relic density \cite{sabrina}. They explain the different behaviour for positive and negative values of $A_0$.

%

\section{Conclusions}

By including the trilinear scalar coupling $A_0$ as a free parameter
in the range of $\pm$4~TeV large regions in the $m_0~-~\m12$ plane of mSUGRA models turn out to be consistent with the WMAP data. Fixing this coupling results in narrow lines strongly depending on $A_0$. Large $m_0$ values (e.g. up to 2 TeV for $\tan\beta~=$~50) are allowed for $|A_0|~\gg$~0. Using the mSUGRA models generated with the ISAJET program we constructed a parametrisation for the common scalar mass $m_0$ as
function of the common gaugino mass $\m12$ for several fixed values of
$A_0$ and $\tan \beta$. The same qualitative behaviour is found using
the SuSpect code. We analysed the discrepancies between the two Monte Carlo programs and traced them back to slightly different mass spectra and couplings, which result in different values of $\Omega_{\chi}h^2$. As the WMAP cut on the relic density is very tight a small difference may have a big impact on the mSUGRA parameter space.\\
In addition we started to study the effect of $A_0$ on the masses of the sparticles. The gluinos and the first two generations of the squarks are insensitive to $A_0$, hence most of the allowed mSUGRA models still lie within the reach of the LHC. However, the third generation of squarks and the heavier Higgs bosons depend significantly on $A_0$.


\subsection*{Acknowledgements}
We would like to thank Michael Spira for helpful theoretical discussions, Luc Pape for useful comments and both for careful reading of this paper. We also thank Josep Flix for providing us with helpful code concerning the linking between the MC programs.



\end{document}